\documentclass[10pt,journal,compsoc]{IEEEtran}

\ifCLASSOPTIONcompsoc
  \usepackage[nocompress]{cite}
\else
  \usepackage{cite}
\fi
\ifCLASSINFOpdf
  \usepackage[pdftex]{graphicx}
  \graphicspath{{../pdf/}{../jpeg/}}
  \DeclareGraphicsExtensions{.pdf,.jpeg,.png,.PNG,.eps}
\else
\fi
\usepackage{url}

\usepackage{array}
\usepackage{multirow}
\usepackage{subcaption}

\begin{document}
%
\title{Vibrotactile Feedback to Make Real Walking in Virtual Reality More Accessible}
%
%
%
%

\author{M. Rasel Mahmud,
        Michael Stewart,
        Alberto Cordova,
        and~John Quarles
\IEEEcompsocitemizethanks{\IEEEcompsocthanksitem M. Rasel Mahmud is with the Department of Computer Science, University of Texas at San Antonio, San Antonio, TX, 78249.\protect\\
E-mail: mrasel.mahmud@my.utsa.edu
\IEEEcompsocthanksitem Michael Stewart is with the Department
of Kinesiology, University of Texas at San Antonio, San Antonio, TX, 78249.\protect\\
E-mail: michael.stewart@utsa.edu
\IEEEcompsocthanksitem Alberto Cordova is with the Department
of Kinesiology, The University of Texas Health Science Center at San Antonio, One UTSA Circle, San Antonio, TX, USA 78229-3901\protect\\
E-mail: alberto.cordova@utsa.edu
\IEEEcompsocthanksitem John Quarles is with the Department
of Computer Science, University of Texas at San Antonio, San Antonio,
TX, 78249.\protect\\
E-mail: john.quarles@utsa.edu.}
\thanks{}}

\IEEEtitleabstractindextext{%
\begin{abstract}
This research aims to examine the effects of various vibrotactile feedback techniques on gait (i.e., walking patterns) in virtual reality (VR). Prior studies have demonstrated that gait disturbances in VR users are significant usability barriers. However, adequate research has not been performed to address this problem. In our study, 39 participants (with mobility impairments: 18, without mobility impairments: 21) performed timed walking tasks in a real-world environment and identical activities in a VR environment with different forms of vibrotactile feedback (spatial, static, and rhythmic). Within-group results revealed that each form of vibrotactile feedback improved gait performance in VR significantly (\textit{p} $<$ .001)  relative to the no vibrotactile condition in VR for individuals with and without mobility impairments. Moreover, spatial vibrotactile feedback increased gait performance significantly (\textit{p} $<$ .001) in both participant groups compared to other vibrotactile conditions. The findings of this research will help to make real walking in VR more accessible for those with and without mobility impairments.


\end{abstract}

}

\maketitle

\IEEEdisplaynontitleabstractindextext

%
\IEEEpeerreviewmaketitle


%
%
%
%

 

\IEEEraisesectionheading{\section{Introduction}\label{sec:introduction}}
\IEEEPARstart{T} he use of virtual reality (VR) is inaccessible to a significant number of individuals with disabilities \cite{agrawal2009disorders,ferdous2016visual,ferdous2018investigating,guo2013effects,samaraweera2013latency}. It is estimated that over one billion, or 15\% of the global population, suffer from a disability \cite{WinNT}. However, research and development in VR rarely consider these individuals, leading to exclusive and inaccessible experiences.  Specifically, many users experience gait disturbances caused by VR, which ultimately limits its usability and benefits \cite{agrawal2009disorders,ferdous2018investigating,guo2013effects}. This issue can be particularly challenging for persons with mobility impairments (MI), as they suffer from functional gait disorders, which make it increasingly difficult for them to use VR technologies. For example, individuals who suffer from MI and other disabilities may find it very challenging to perform various locomotor movements in VR without fear of falling or injuring themselves. Despite these challenges, little research has been conducted in order to mitigate them.
 Research in the field of assistive technologies has revealed that some multimodal feedback techniques \cite{franco2012ibalance,sienko2017role} can assist individuals with MI with their daily activities as well as improve gait and balance. A number of studies have been conducted with the aim of improving balance and gait for persons with disabilities using assistive technology based on visual feedback \cite{velazquez2010wearable,thikey2011need,vcakrt2010exercise,sutbeyaz2007mirror}. However, one of the major issues with visual feedback is that it interferes with the presence in VR more than the vibrotactile feedback \cite{gibbs2022comparison}. Nevertheless, vibrotactile feedback to improve gait and balance in VR has received very little attention. Our study investigated the mitigation of gait disturbance issues by implementing various vibrotactile feedback techniques (e.g., spatial, static, and rhythmic) in VR for participants with and without MI. We conducted a timed walking task using a pressure-sensitive walkway (GAITRite) for the purpose of quantitative gait analysis. However, these feedback effects on gait performance were not measured post-study. The purpose of this study was to make real walking in immersive VR more accessible by using vibrotactile feedback, as well as to examine its effectiveness in enhancing gait efficiency. 

\section{BACKGROUND AND RELATED WORK}
\label{sec:related-work}

\subsection{Gait Disturbances in VR}
The use of VR has been shown to cause instability and gait disturbances in previous studies. Additionally, the use of head-mounted displays (HMDs) can result in individuals losing stability as a result of end-to-end latency and illusory impressions of their body movements. HMDs block visuals from the real world due to the obstructing of visual feedback \cite{soltani2020influence, martinez2018analysing}. Additionally, long-term exposure to VR caused postural instability \cite{murata2004effects}. As a result of these postural instabilities, walking in a virtual environment (VE) can lead to gait instability \cite{hollman2007does}. In addition, Riem et al. \cite{riem2020effect} found that step lengths were significantly affected by VR in comparison to baseline settings (\textit{p} $<$ .05). In other studies \cite{sondell2005altered}, imbalance and gait disturbance in VR have also been reported. A study conducted by Horsak et al. \cite{horsak2021overground} examined the gait differences of 21 participants (male: 9, female:12, age: 37.62 ± 8.55 years) during walking in an HMD-based VE. Their findings revealed that walking speed was reduced by 7.3\% in the HMD-based VE. A study by Canessa et al. analyzed the differences between real-world walking and immersive VR walking while wearing an HMD \cite{canessa2019comparing}. Based on their results, they found that walking velocity decreased significantly (\textit{p} $<$ .05) in immersive virtual reality when compared to the real world. A VR HMD with continuous multidirectional visual field perturbation was used by Martelli et al. \cite{martelli2019gait} in order to study how healthy young adults' gaits were modified and altered as they walked in a VE while using a VR Headset. Twelve healthy young adults walked for six minutes on a pathway in four different settings. As a result of perturbation of the visual field, stride length, width, and variability of stride were reduced. Despite these gait disturbance issues in VR, few attempts have been made in the past to address it. Thus, we investigated these gait disturbance issues in order to ease the experiences of walking in VR and make VR more accessible.

\subsection{Gait Improvement After VR Intervention} Despite the focus of our research on VR accessibility and gait improvement while in VR, it is important to examine how VR rehabilitation applications have previously been used to foster balance and gait improvement that persists after the VR experience has ended \cite{de2016effect,meldrum2012effectiveness,park2015effects,cho2016treadmill,duque2013effects, bergeron2015use}. 

A study by Walker et al. \cite{walker2010virtual} investigated the effectiveness of a low-cost VR system in improving walking and balance abilities among seven post-stroke patients. By displaying a television screen in front of a treadmill, participants were able to experience walking along a city street. Head-mounted position sensors were used to collect postural feedback. During the study, all participants were supported by an overhead suspension harness. In the study, participants were within a year after stroke and had previously received traditional rehabilitation but had significant gait problems as well. The study included results from six participants (mean age 53.5 years, range 49-62 years) and excluded one participant due to sickness. According to the results of the study, there was a significant improvement (\textit{p} $<$ .05) in balance, walking speed, and gait functionality after the study. They reported 10\% improvements in Berg Balance Scale (BBS), 38\% improvements in walking speed, and 30\% improvements in Functional Gait Assessment (FGA) scores. 

In a study conducted by Janeh et al. \cite{janeh2019gait}, 15 male patients with Parkinson's disease were recruited to investigate the effectiveness of a VR-based gait manipulation approach aimed at adjusting step length in order to achieve gait symmetry. By using visual and proprioceptive signals, they were able to compare natural gait with walking situations when performing VR-based gait activities. In comparison to natural gait, VR gait activities increased step width and swing time. Furthermore, Janeh et al. reported that VR might improve the gait of persons with neurological disorders after experiencing it. As a result of their observations, they stressed the importance of using virtual walking approaches in rehabilitation \cite{janeh2021review}.

Nevertheless, the majority of gait rehabilitation approaches did not use VR techniques. In our study, we investigated immersive VR-based walking where the VEs were rendered using HMDs.

\subsection{Gait Disturbances Associated With HMDs for Participants With MI}
A study by Winter et al. \cite{winter2021immersive} recruited 36 participants (Male: 10, Female: 26) without a history of MI and 14 patients (MS: 10, Stroke: 4) with MI to investigate the effect of an immersive, semi-immersive, and no VR environment on gait during treadmill training. Participants first completed the treadmill training without VR. In the semi-immersive VR condition, participants were presented with a virtual walking path on a monitor. In the immersive VR condition, the participants experienced the same VR scenario via HMDs. The results of the study indicated that immersive VR during gait rehabilitation increased walking speed more significantly (\textit{p} $<$ .001) than semi-immersive and no VR conditions for both participant groups. VR conditions did not cause cybersickness or an increase in heart rate.

In addition, Guo et al. \cite{guo2015mobility} examined the effect of VEs on gait in both participants with and without MI. It was found that MI participants responded differently in terms of walking speed, step length, and stride length when compared to non-MI participants. Despite this, other gait parameters did not differ significantly between participants with and without MI. 

Ferdous et al. \cite{ferdous2018investigating} analyzed the effects of HMDs and visual components on postural stability in VR for participants with multiple sclerosis (MS) using visual feedback. However, they did not examine the effect of vibrotactile feedback on gait, as in most previous studies on immersive VR with HMDs. Therefore, the impact of vibrotactile feedback on gait in immersive VR with HMDs has received insufficient attention. Also, most prior studies have primarily focused on individuals without MI \cite{lott2003effect,epure2014effect,robert2016effect,horlings2009influence,samaraweera2015applying}. These led us to examine the vibrotactile feedback effect on gait for participants with and without MI using immersive VR with HMDs.

\subsection{Vibrotactile Feedback for Gait Improvement in the Real World }
Prior studies reported that visual feedback \cite{alahakone2010real}, auditory feedback \cite{chiari2005audio}, or vibrotactile feedback \cite{sienko2012biofeedback} in real-world applications reduced postural instability and gait disturbances for participants with balance and gait impairments \cite{henry2019age, wannstedt1978use, sienko2018potential}. Vibrotactile feedback is generally favored over other helpful feedback modalities because it is considered to interact less with other senses, such as seeing or hearing, which may be restricted by visual or auditory feedback \cite{goodworth2009influence, wall2009vibrotactile}. A few research have studied the use of vibrotactile feedback to enhance balance and gait in real-world applications (e.g., rehabilitation).

For example, Rust et al. examined the impact of vibrotactile feedback on trunk sway in fifteen MS patients in the real world \cite{rust2020benefits}. The participants wore a headband containing eight 150 Hz vibrators spaced at 45-degree intervals. Vibrators were turned on when a sway threshold in the vibrator's direction was exceeded. In four weeks, participants performed a variety of training, gait, and balancing activities. The authors initially assessed trunk sway as a baseline. Using the SwayStar device, they detected trunk sway with vibrotactile input. After one and two weeks of training with vibrotactile feedback, there was a large decrease in trunk sway (\textit{p} $<$.02) compared to the baseline. The authors measured a carry-over effect in the fourth week following three weeks of no training. In addition, they discovered a considerable carry-over improvement (\textit{p} $<$.02). In their experiment, standing with eyes closed on a foam pad had the most significant results, with a 59\% reduction in pitch sway (\textit{p} $<$.002).

Ballardini et al. \cite{ballardini2020vibrotactile} recruited 24 participants (11 males, 13 females) to examine the influence of vibrotactile feedback on standing balance and gait performance. They designed a system that provides vibrotactile feedback using two vibration motors located on the front and rear of the body. An accelerometric measurement encoding that combines the position and acceleration of the body in the anterior-posterior direction was synced with the vibration. The objective was to examine two different encoding techniques: 1) constant vibration and 2) vibration with a dead zone (i.e., silence when the signal was below the given threshold). Using vibrations unrelated to the actual postural oscillations, they determined whether the informative quality of the input altered these effects (sham feedback). Nine participants experienced the vibration always on and sham feedback, while fifteen received vibration with a dead zone. According to the results, synchronized vibrotactile feedback lowered postural sway significantly in the anterior-posterior and medial-lateral directions.  In terms of reducing postural sway, there was no significant difference between the two encoding methods. The presence of sham vibration feedback enhanced postural sway, highlighting the significance of the encoded data.

Thirty-nine participants with an imbalance and mobility issues due to the severe bilateral vestibular loss were recruited by Kingma et al. to examine how vibrotactile feedback impacts balance and movement in the real world \cite{kingma2019vibrotactile}. The participants wore a vibrotactile belt around their waists for two hours each day for one month. If they were in motion, they had to wear the belt. The belt's 12 motors were actuated by a microprocessor. Before and after one month of daily usage of the belt, participants were asked to verbally assess their balance and mobility on a 0 to 10 scale. The average ratings for mobility and balance grew significantly (\textit{p} $<$.00001) when compared to those obtained without the belt.

While these studies explored balance and gait in the real world, we investigated the effects of various vibrotactile feedback on gait in VR for both participants with and without MI.

\section{HYPOTHESES}
\label{sec:hypotheses}

This research investigated the effects of three vibrotactile feedback conditions (spatial, static, and rhythmic) on gait in a VR environment. 
We were inspired by three types of audio feedback that were proven to be successful in VR \cite{mahmud2022auditory} and non-VR contexts \cite{stevens2016auditory,gandemer2017spatial, ross2016auditory,cornwell2020walking,ghai2018effect}. Further, based on previous research on vibrotactile feedback for assistive technology, the following hypotheses were investigated:

H1: Compared to non-VR baseline without vibrotactile feedback condition, VR baseline without vibrotactile feedback will exhibit gait disturbances.

H2: Three VR-based vibrotactile feedback conditions (spatial, static, and rhythmic) will improve gait metrics more than the condition with no vibrotactile feedback in VR.

H3: Spatial vibrotactile feedback will enhance gait metrics more than static and rhythmic vibrotactile feedback.

H4: While experiencing the vibrotactile feedback in VR, participants with MI will encounter greater gait improvement (e.g., velocity) compared to participants without MI.
\section{METHODS}

\subsection{Participants, Selection Criteria, and Screening Process}
In order to investigate gait improvement using vibrotactile feedback in VR, 39 participants (male: 18, female: 21) were recruited from multiple sclerosis support groups and the local community. In this study, 18 participants (male: 6, female: 12) had MI due to MS. Among the participants with MI, 44.44\% were White, 25.56\% were Hispanic, 4.44\% were Asian, and 25.56\% were African American. Furthermore, we recruited a group of twenty-one participants without MI (male: 11; female: 10). Participants without MI constituted 18.18\% White, 22.73\% Hispanic, 50\% African American, 4.55\% American Indian, and 4.54\% Asian. Descriptive statistics (mean and standard deviation of age, height, weight, etc.) of participants in both groups can be found in Table 1. Exclusion criteria for participants included those with cognitive impairments, severe vision impairments, cardiovascular or respiratory conditions, or those unable to walk without assistance. We intentionally recruited people with MI due to MS because this population is rarely considered during VR development, which may lead to inaccessible and non-inclusive experiences.

\begin{table}[ht!]
\caption{Descriptive statistics for participants}
    \label{tab:my_label}
\setlength{\tabcolsep}{2 pt}
\begin{tabular}{|c|cc|cc|cc|cc|}
\hline
\multirow{0}{*}{}{\textbf{\begin{tabular}[c]{@{}c@{}}Participant \\ Group\end{tabular}}} & \multicolumn{2}{c|}{\textbf{Participants}} & \multicolumn{2}{c|}{\textbf{Age (years)}} & \multicolumn{2}{c|}{\textbf{Height (cm)}} & \multicolumn{2}{c|}{\textbf{Weight (kg)}} \\ \cline{2-9} 
                                                                                       & \multicolumn{1}{c|}{Male}     & Female     & \multicolumn{1}{c|}{Mean}      & SD       & \multicolumn{1}{c|}{Mean}      & SD       & \multicolumn{1}{c|}{Mean}      & SD       \\ \hline
\textbf{MI}                                                                            & \multicolumn{1}{c|}{6}        & 12         & \multicolumn{1}{c|}{45.89}      & 10.18     & \multicolumn{1}{c|}{165.60}    & 10.26    & \multicolumn{1}{c|}{83.86}     & 28.27    \\ \hline
\textbf{Without MI}                                                                    & \multicolumn{1}{c|}{11}        & 10         & \multicolumn{1}{c|}{47.29}      & 12.09     & \multicolumn{1}{c|}{166.12}    & 10.05     & \multicolumn{1}{c|}{88.26}     & 16.05    \\ \hline
\end{tabular}
\end{table}

Screening Process:
Several methods were used to recruit participants, including telephone calls, email lists, and flyers. The eligibility of participants was determined via a telephone pre-screening process. In order to determine whether a participant could be included or excluded from the study, we asked about their general demographics, health, and medical histories. As an example, we confirmed that the individual could visit the on-campus lab and participate throughout the study. Additionally, we assessed the participants' history of MI and their ability to walk independently. In order to minimize participant characteristic imbalances, we ensured that the age, height, and weight of both participants were proportionally the same (see Table 1).

\subsection{System Description}
The study used a variety of equipment to ensure the safety of participants and collect data.

\subsubsection{Vibrotactile Equipment:}					
We used the following vibrotactile equipment from bHaptics (https://www.bhaptics.com):	

Vest: Participants wore a wireless vest with 40 independently controlled Eccentric Rotating Mass (ERM) vibrotactile motors. Twenty vibrotactile motors were placed anteriorly on the vest, and the remaining twenty were placed posteriorly. The vest was fitted with shoulder snap buttons that allowed it to be adjusted. The total weight of the vest was 1.6 kg.

Arm Sleeves: The participants wore arm sleeves located between the wrist and elbow that had adjustable straps. The arm pieces contained six ERM vibration motors. The arm pieces weighed 0.30 kg each.

Forehead: Six ERM vibrotactile motors were attached to the HMD and were positioned on each participant's forehead. It had a weight of 0.08 kg. 

Fig. 1 depicts the placement of the vibration motors.

\begin{figure}[h!]
\centering
\includegraphics[width=8.25cm,height=5cm]{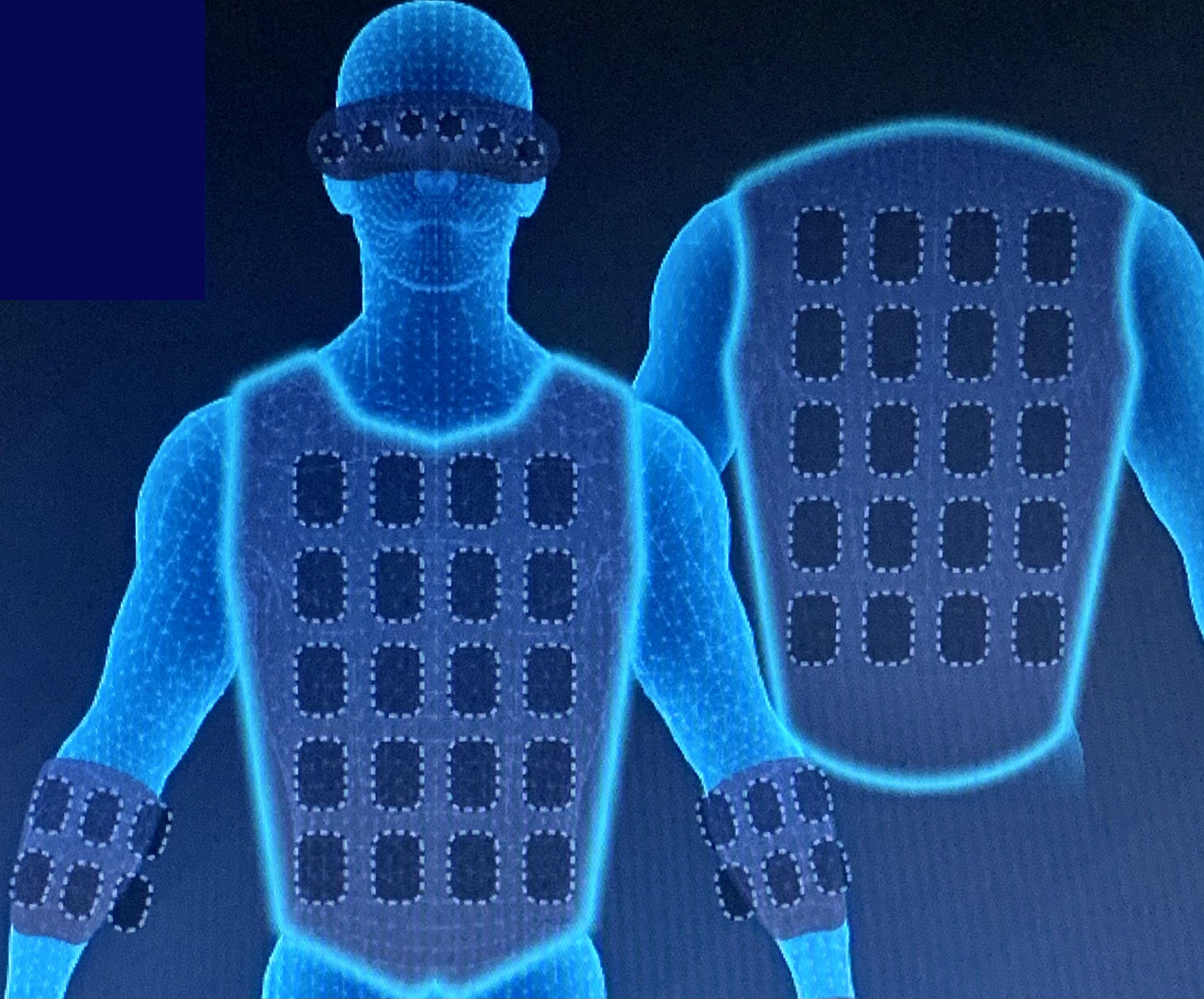}
\caption{ Vibration motors positions.}
\end{figure}

\subsubsection{Computers, VR Equipment, and Software:}
The VEs were developed using Unity3D software. The HTC Vive wireless HMD was used for our experiment, which had a pixel resolution of 2160 x 1200, a 90 Hz refresh rate, and a 110-degree field of view. To track the position of the bHaptics vest, we installed a Vive tracker on the back of the vest. A computer with an Intel Core i7 processor (4.20 GHz), 32 GB DDR3 RAM, and an NVIDIA GeForce RTX 2080 graphics card was used to render the VE and record the data.

\subsubsection{Audio-to-Vibrotactile Software:} 
A vibrotactile output is generated from corresponding audio input with this software. First, we attached the audio input to the Unity scenes. Then with the help of the bHaptics Unity plugin, we delivered the audio input to bHaptics audio-to-vibrotactile software, which converted the audio input to the corresponding vibration. White noise was used to generate audio input rather than music or other user-selected tones as white noise has been shown to change the signal-to-noise ratio and increase balance performance due to the stochastic resonance phenomenon \cite{helps2014different}. Furthermore, bHaptics software allowed us to control the intensity of the vibrations. We adjusted vibration intensity until participants stated it was a comfortable intensity.

\subsubsection{Safety Equipment:} 
For the safety of our participants, we used a Kaye Products Inc. suspension walking system consisting of a body harness, thigh cuffs, and suspension walker.

\subsubsection{Gait Analysis:}
We collected gait metrics from participants using the GAITRite walkway system. The system consists of a portable 12 feet pressure sensor pad capable of measuring the gait metrics of participants during a walking test. The GAITRite walkway system has the capability to provide spatial and temporal gait data of the participants.
\subsubsection{Environment:} 
A controlled lab environment was used during the study ($>$600 square feet.). The participants and researcher were the only individuals present in the room in order to minimize any noise or other disruptions from the surrounding environment. A comparison of the real-world and virtual environments for the timed walking task can be seen in Fig. 2.

\begin{figure}[h!]
    \centering
    \includegraphics[width=0.20\textwidth, height=8.99cm, angle=270]{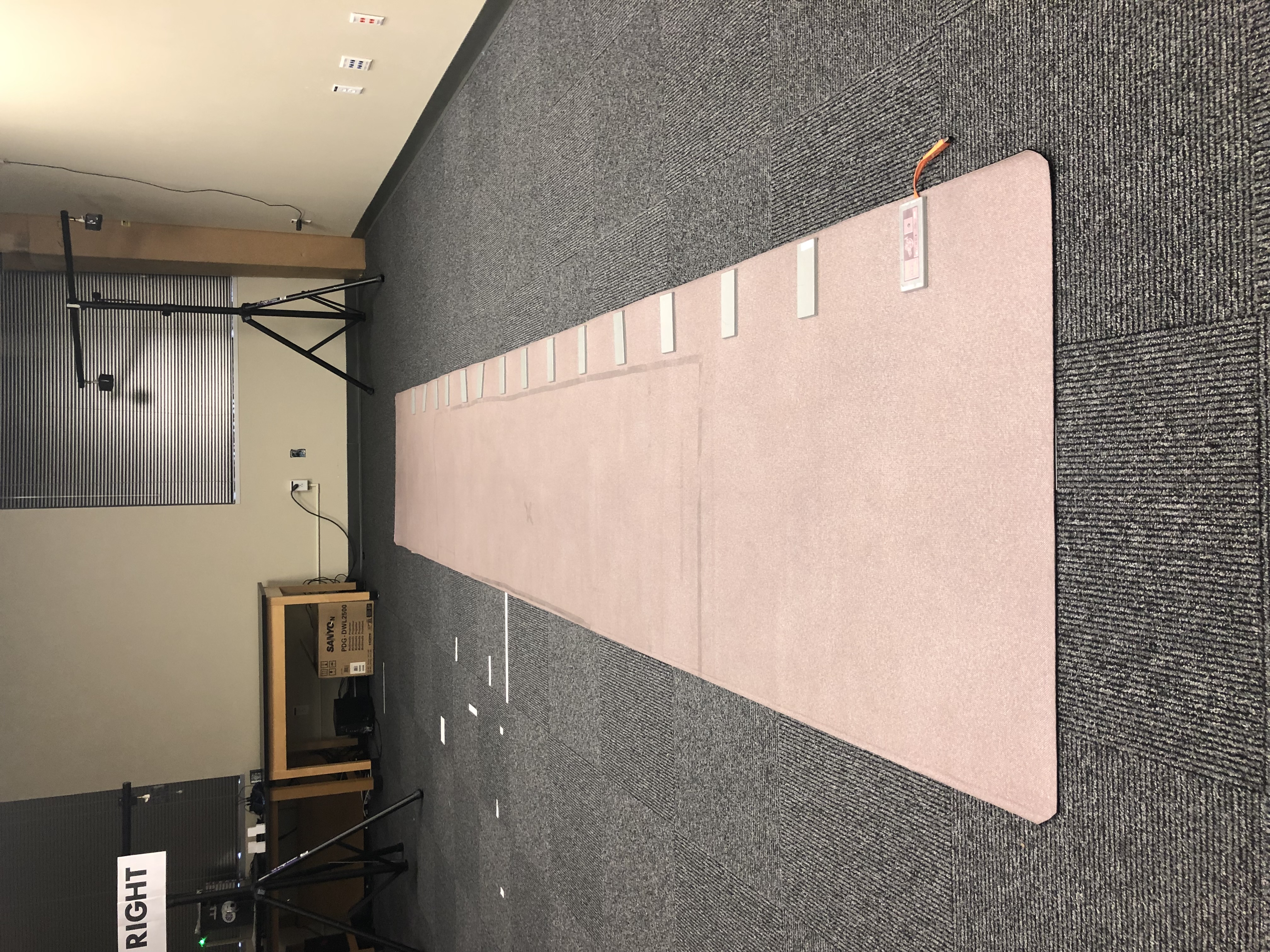}
  \includegraphics[width=0.494\textwidth,height=4cm]{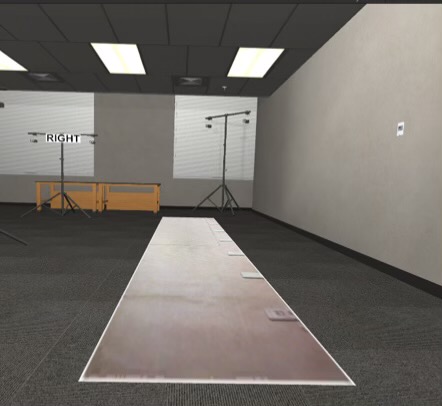}
  \caption{ Comparison between real environment (top) and virtual environment (bottom) for timed walking task.}
\end{figure}

\subsection{Study Conditions}
We evaluated three kinds of VR-based vibrotactile feedback techniques and a condition without vibrotactile feedback to assess how vibrotactile feedback affects gait in VR. The audio-to-vibrotactile software from bHaptics converted the audio into corresponding vibrotactile feedback for each feedback condition (see section 4.2.3).


\subsubsection{Non-VR Baseline Trials}
Participants performed the timed walking task using the GAITRite system without VR and without receiving any vibrotactile feedback to record participants’ baseline gait metrics.

\subsubsection{VR Baseline Trials}
A VR baseline measurement for participants was established by performing the same timed walking task without the application of any vibrotactile feedback in VR. HMDs and the bHaptics equipment pieces were worn by participants during this condition.

\subsubsection{Spatial Vibrotactile Feedback} 
To generate spatial vibrotactile feedback, the audio was used to set the vibrotactile feedback patterns. We used Google resonance audio SDK in Unity for audio spatialization since the plugin uses head-related transfer functions (HRTFs) to replicate 3D sound more precisely than Unity's default \cite{chong2020audio, pinkl2020spatialized}. The spatial audio in our research was simulated \cite{kim2019immersive} rather than recorded ambisonic audio \cite{mccormack2022parametric}. The spatialized audio simulation was then sent to the audio-to-vibrotactile software, which created spatial vibrotactile feedback. The forehead bHaptics equipment vibrated to varying degrees as the user rotated their head. The vibration of the vest was altered based on its position as sensed by the Vive tracker. The X, Y, and Z coordinates of the 3D audio source and participant in the VE were 0, 1, 0, and 0, 0, 0, respectively, indicating that the audio source was positioned directly in front of the participant.

\subsubsection{Static Vibrotactile Feedback}  
We transmitted white noise to the audio-to-vibrotactile software of bHaptics in order to provide static vibrotactile feedback. All the vibration motors in the forehead, arm sleeves, and vest were vibrated continuously. The user's location had no influence on the feedback. This technique has been shown in non-VR studies to improve the balance of adults \cite{ross2016auditory}.

\subsubsection{Rhythmic Vibrotactle Feedback} 
We transmitted a white noise beat at 1-second intervals to the bHaptics audio-to-vibrotactile software in order to generate rhythmic vibrotactile feedback. All the vibration motors in the forehead, arm sleeves, and vest were vibrated. The length of the rhythmic feedback beat was also 1 second. The previous study has shown that experiencing a steady beat may improve balance and gait in both those with neurological impairments and older people in non-VR environments \cite{ghai2018effect}.

\subsubsection{No Vibrotactile Feedback} 
This was used to measure participants' balance in VR without vibrotactile input. In order to remain coherent with previous conditions, participants continued to wear the HMD, bHaptics suit, arm sleeves, and forehead component, but no vibrotactile input was supplied.

\subsection{Study Procedures}
Fig. 3 is the flowchart of the whole study procedure.

\begin{figure}[ht!]
    \centering
  \includegraphics[width=0.47\textwidth,height=10cm]{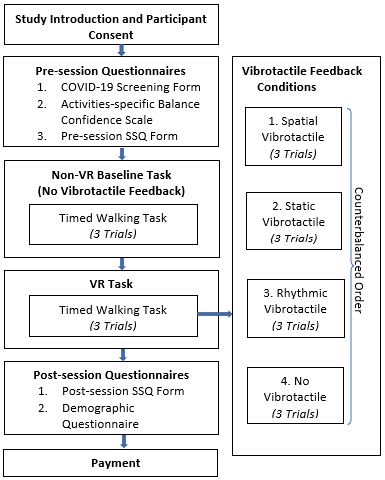}
  \caption{Study procedure}
\end{figure}

We first sanitized all the lab equipment (including the HMD, Vive tracker, bHaptics devices, safety harness, and suspension system). A COVID-19 symptom screening questionnaire was completed when participants entered the lab. The research procedures were then explained to the participants, and formal consent was documented. 

\subsubsection{Pre-Study Questionnaires}There were two questionnaires completed at the beginning of the study: an Activities-specific Balance Confidence (ABC) form \cite{powell1995activities} and an SSQ questionnaire \cite{kennedy1993simulator}. We requested that participants take off any footwear that could interfere with the GAITRite system. Study approval was granted by the Institutional Review Board (IRB).

\subsubsection{Real World Walking}
A GAITRite walkway was used to measure gait metrics in this study. To prevent fall-related injuries, participants were securely fastened to safety harnesses and suspension walker. Participants were instructed to walk on the GAITRite at a speed that they found comfortable. In addition, we instructed them to complete 180-degree turns at both ends. The GAITRite software requires participants to step off between trials, as the system cannot accurately assess turns. Three timed walking trials were conducted \cite{steffen2002age} while a stopwatch was used to measure the walking time of the participants.

\begin{figure}[ht!]
	\centering
	\includegraphics[width=1.00\linewidth]{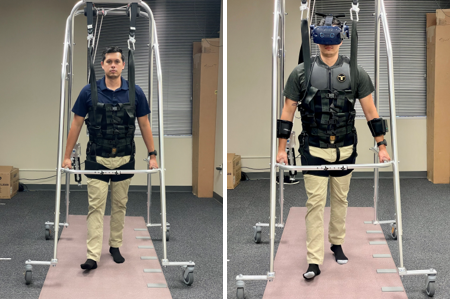}
	\caption{Participants used harness while performing the timed walking task utilizing the GAITRite system in both the real world (left) and virtual environment (right).}
\end{figure}

\subsubsection{Virtual Environment Walking}
We replicated the walking task in a VE with various vibrotactile feedback conditions in order to simulate the real-world walking environment. We used the same harness and suspension system as in the real-world environment to prevent sudden falls. During the study, participants were instructed to walk on the virtual GAITRite, which was layered over the physical GAITRite. HMDs were used to observe the VE while the vibrotactile feedback was applied through the bHaptics equipment. Three timed walking trials were performed in VR for each vibrotactile condition (e.g., spatial, static, and rhythmic) and a no-vibrotactile in VR condition. A counterbalanced order of the four conditions was applied to all participants in the study. Fig. 4 shows the comparison of real and virtual environment walking for the participants.

\subsubsection{Post-Study Questionnaires}
A post-study SSQ questionnaire and a demographic form were completed by the participants at the end of the study. Finally, each participant received \$30/hour compensation and a parking validation ticket.

\section{METRICS}
\label{sec:metrics}

\subsection{Gait Metrics}
We examined the following gait metrics relative to each vibrotactile feedback condition in this study.\\
\textit{- Walking Velocity}: The distance traveled (cm) divided by ambulation time (sec).\\
\textit{- Cadence}: The number of steps taken per minute.\\
\textit{ - Step Time (Left/Right)}: The amount of time (sec) between the initial contact points of the opposite foot.\\ 
\textit{- Step Length (Left/Right)}: The distance (cm) between the centers of the heels of two successive steps taken by opposing feet.\\
\textit{- Cycle Time (Left/Right)}: The time (sec) between the initial contact points of the same foot's two successive steps.\\
\textit{- Stride Length (Left/Right)}: The distance (cm) between the steps of the same foot.\\
\textit{- Swing Time (Left/Right)}: The time (sec) between a foot's ultimate contact point and its starting contact point.\\
\textit{- Stance Time (Left/Right)}: The time (sec) between the start and final contact points of a single footstep.\\
\textit{- Single Support Time (Left/Right)}: This is the time (sec) between the final contact of the current footfall and the first contact of the following footfall of the same foot.\\
\textit{- Double Support Time (Left/Right)}: The time (sec) when both feet are on the ground.\\
\textit{- Base of Support (Left/Right)}: The width between one foot and the progression line of the opposing footstep. \\
\textit{- Toe-In/Toe-Out (Left/Right)}: The angle (degrees) between the progression line and footprint's midline. Toe-in means that the center-line of the footprint is inside the line of progression. Toe-out denotes that the center-line of the footprint is outside the line of progression.\\
The GAITRite manual has further details on Gait metrics \cite{WinNT3}.

\subsection{Activities-specific Balance Confidence (ABC) Scale}
Participants were assessed on their balance, mobility, and physical functionality using the Activities-specific Balance Confidence Scale (ABC). Sixteen items are used in this questionnaire to determine whether an individual is confident in performing daily functions without losing balance \cite{powell1995activities}. Each participant is asked to rate their level of confidence in each specific activity on a scale of 0\% (not confident) to 100\% (most confident). ABC Scale scores are calculated by dividing the sum of the ratings (0-1600) by 16. ABC scores below 50 indicate low functioning. Additionally, ABC scores between 50-80 indicate moderate levels of functioning, and ABC scores above 80 indicate high levels of functioning.

\subsection{Simulator Sickness Questionnaire (SSQ)}
Cybersickness of the participants resulting from virtual environment exposure was measured by the Simulator Sickness Questionnaire (SSQ). The SSQ assesses participants' physiological discomfort due to cybersickness using 16 symptoms that are organized into three different categories (disorientation, nausea, and oculomotor disturbance) \cite{kennedy1993simulator}. 
\section{STATISTICAL ANALYSIS}
\label{sec:statistical}

For each investigated gait metric, the Shapiro-Wilk test (p $>$ .05), histograms, normal Q-Q plots, and box plots revealed that the data was normally distributed for both participants with and without MI. Then, we performed a 2$\times$5 mixed-model ANOVA with two between-subject factors (participants with MI and participants without MI) and five within-subject factors (five study conditions: baseline, spatial, static, rhythmic, and no vibrotactile) to identify any significant difference in study conditions. When there was a significant difference, post-hoc two-tailed t-tests were performed for pairwise within and between-group comparisons to find the specific differences between two study conditions. For cybersickness analysis, we also used two-tailed t-tests comparing pre-session and post-session SSQ scores for each participant group. Additionally, we conducted two-tailed t-tests comparing the ABC scores of both participant groups to assess the difference in physical ability. Bonferroni corrections were applied to all tests involving multiple comparisons.
\section{RESULTS}
\label{sec:results}
Seven of the twelve studied gait metrics (walking velocity, cadence, step length, stride length, step time, cycle time, and swing time) improved significantly under varied vibrotactile feedback conditions, but the other five gait metrics improved insignificantly for both groups of participants. Gait improvement also varied greatly depending on the different vibrotactile feedback conditions. We assessed the gait characteristics from the start to finish of the trials (not any specific portion of the trials). Additionally, both left and right leg data were evaluated. There was no significant difference in the results for the left and right legs data. For simplicity, we presented the average data for the left and right leg for all gait metrics.

The mixed-model ANOVA test for all individuals revealed a significant difference in walking velocity, \textit{F}(1,123) = 69.8, \textit{p} $<$ .001; and effect size, $\eta^{2}$ = 0.08. In addition, we discovered a statistically significant difference (\textit{p} $<$ .001) in cadence, step length, stride length, step time, cycle time, and swing time. Then we did post-hoc paired t-tests for within-group and two-tailed t-tests for between-group comparisons to identify differences between specific research conditions.

\subsection{Participants With MI: Within-Group Comparisons}

\subsubsection{Non-VR Baseline vs. VR Baseline}
Walking velocity was significantly lower in the VR baseline without vibrotactile feedback condition (Mean, \textit{M} = 96.69, Standard Deviation, \textit{SD} = 13) compared to non-VR baseline without vibrotactile feedback condition (\textit{M} = 128.25, \textit{SD} = 10.84); \textit{t}(17) = 8.19, \textit{p} $<$ .001; and effect size, Cohen's \textit{d} = 0.11. We also found a significant reduction (\textit{p} $<$ .001) in cadence, step length, and stride length for the VR baseline without vibrotactile feedback condition. Step time, cycle time, and swing time were significantly enhanced (\textit{p} $<$ .001) in VR baseline without vibrotactile feedback condition than non-VR baseline without vibrotactile feedback condition. These findings suggested that participants with MI experienced gait disturbances in VR environments.
 
\subsubsection{Spatial Vibrotactiile vs. VR Baseline}
Walking velocity enhanced significantly in the spatial vibrotactile feedback condition (\textit{M} = 127.35, \textit{SD} = 15.03) compared to VR baseline without vibrotactile feedback condition (\textit{M} = 96.69, \textit{SD} = 13); \textit{t}(17) = 5.74, \textit{p} $<$ .001, \textit{d} = 0.76. We found a significant increase (\textit{p} $<$ .001) in cadence, step length, and stride length for spatial vibrotactile feedback condition compared to VR baseline without vibrotactile feedback condition. Results also demonstrated a significant reduction in step time, cycle time, and swing time (\textit{p} $<$ .001) in spatial vibrotactile feedback condition compared to VR baseline without vibrotactile feedback condition. The findings demonstrated that spatial vibrotactile feedback enhanced gait metrics compared to the VR baseline without vibrotactile feedback condition.

\begin{figure}[ht!]
    \centering
  \includegraphics[width=0.5\textwidth,height=6 cm]{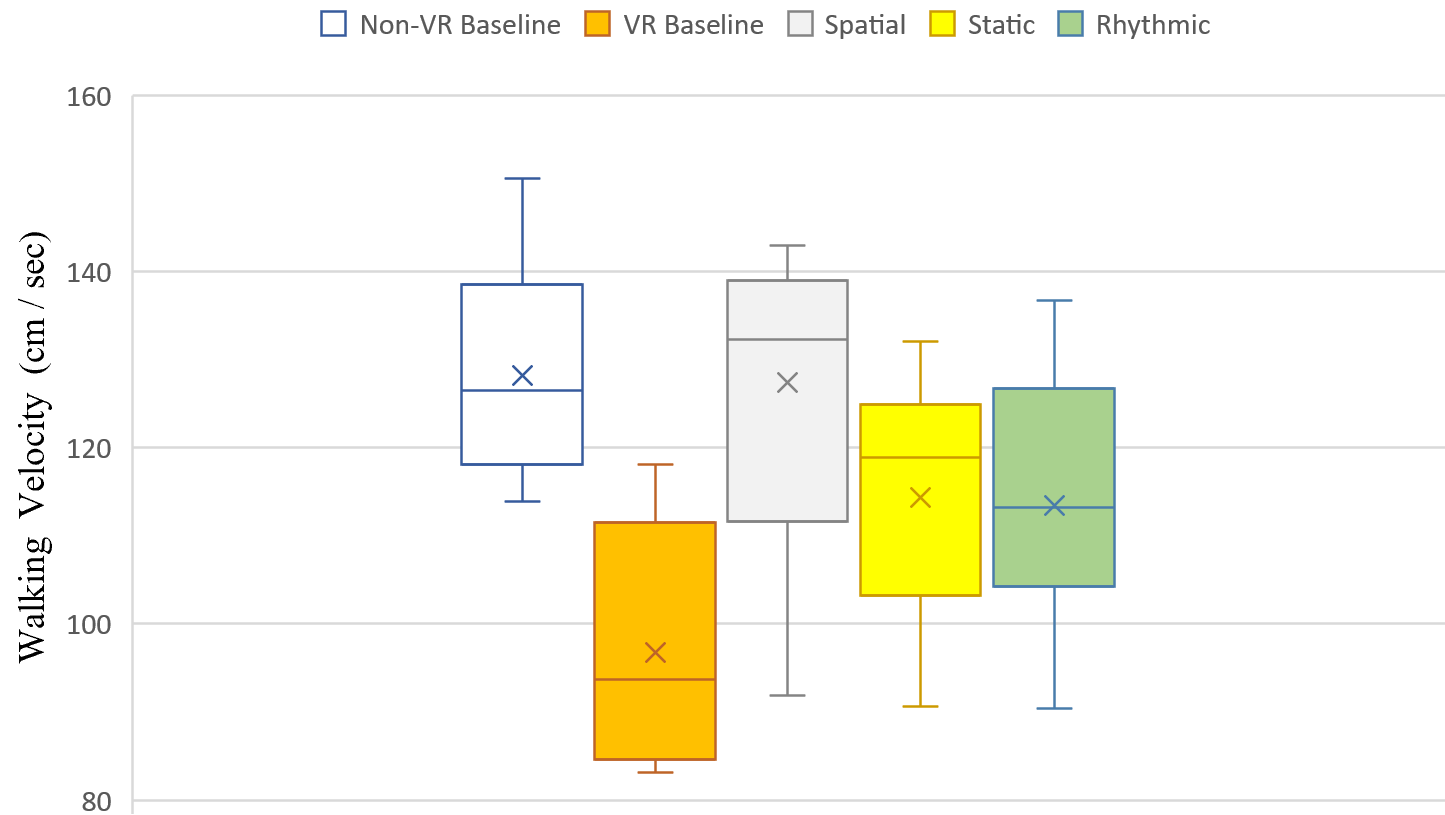}
   \caption{ A comparison of walking velocity between study conditions for participants with MI.}
\end{figure}

\subsubsection{Spatial Vibrotactile vs. Static Vibrotactile}
Experimental results revealed that walking velocity was increased in spatial vibrotactile feedback condition (\textit{M} = 127.35, \textit{SD} = 15.03) relative to static vibrotactile feedback condition (\textit{M} = 114.4, \textit{SD} = 12.6); \textit{t}(17) = 4.71, \textit{p} $<$ .001, \textit{d} = 0.27. We discovered a statistically significant increase (\textit{p} $<$ .001) in cadence, step length, and stride length for spatial vibrotactile feedback condition than static vibrotactile feedback condition. Additionally, we noticed that step time, cycle time, and swing time were significantly reduced (\textit{p} $<$ .001) in the spatial vibrotactile feedback condition compared to the the static vibrotactile condition. Therefore, spatial vibrotactile feedback condition exhibited greater gait performance compared to the static vibrotactile condition. 

\subsubsection{Spatial Vibrotactile vs. Rhythmic Vibrotactile}
We noticed a significant increase in walking velocity in spatial (\textit{M} = 127.35, \textit{SD} = 15.03) compared to rhythmic vibrotactile feedback (\textit{M} = 113.45, \textit{SD} = 13.68); \textit{t}(17) = 4.1, \textit{p} $<$ .001, \textit{d} = 0.36. Also, cadence, step length, and stride length for spatial vibrotactile feedback increased significantly (\textit{p} $<$ .001) than the rhythmic vibrotactile feedback. However, we found a significant reduction (\textit{p} $<$ .001) in step time, cycle time, and swing time in spatial compared to rhythmic vibrotactile feedback. The findings showed that spatial vibrotactile feedback might be more beneficial for gait performance than rhythmic vibrotactile feedback.

\subsubsection{Static Vibrotactile vs. VR Baseline}
Our results indicated that walking velocity was significantly increased in static vibrotactile (\textit{M} = 114.4, \textit{SD} = 12.6) compared to VR baseline without vibrotactile feedback (\textit{M} = 96.69, \textit{SD} = 13); \textit{t}(17) = 4.84, \textit{p} $<$ .001, \textit{d} = 0.58. We  also found a significant increase (\textit{p} $<$ .001) in cadence, step length, and stride length in static than VR baseline without vibrotactile feedback. However, we also noticed a  significant reduction (\textit{p} $<$ .001) in step time, cycle time, and swing time in static than VR baseline without vibrotactile feedback. Thus, static vibrotactile feedback outperformed VR baseline without vibrotactile feedback.

\subsubsection{Static Vibrotactile vs. Rhythmic Vibrotactile}
We did not find a significant difference in walking velocity between static (\textit{M} = 114.4, \textit{SD} = 12.6) and rhythmic vibrotactile feedback (\textit{M} = 113.45, \textit{SD} = 13.68); \textit{t}(17) = 0.35, \textit{p} $=$ .73, \textit{d} = 0.12 after post-hoc two-tailed paired t-test. We also noticed no significant difference for other gait parameters between static and rhythmic vibrotactile feedback condition. Therefore, study was equivocal as to whether rhythmic or static vibrotactile input is more efficient for improving gait performance.

\begin{table}[ht!]
\caption{Gait metrics in five conditions for participants with MI}
    \label{tab:my_label}
\begin{tabular}{|l|l|l|l|l|l|}
\hline
\begin{tabular}[c]{@{}l@{}}Gait \\ Metrics\end{tabular}           & \begin{tabular}[c]{@{}l@{}}Non-VR \\ baseline\\ \\ Mean\\ (SD)\end{tabular} & \begin{tabular}[c]{@{}l@{}}VR\\ baseline \\ \\ Mean\\ (SD)\end{tabular} & \begin{tabular}[c]{@{}l@{}}Spatial \\ \\ Mean\\ (SD)\end{tabular} & \begin{tabular}[c]{@{}l@{}}Rhythmic\\ \\ Mean\\ (SD)\end{tabular} & \begin{tabular}[c]{@{}l@{}}Static   \\ \\ Mean\\ (SD)\end{tabular} \\ \hline
Cadence                                                           & \begin{tabular}[c]{@{}l@{}}113.67\\ (9.84)\end{tabular}                     & \begin{tabular}[c]{@{}l@{}}92.14\\ (8.56)\end{tabular}                     & \begin{tabular}[c]{@{}l@{}}111.33\\ (6.17)\end{tabular}          & \begin{tabular}[c]{@{}l@{}}103.65\\ (5.03)\end{tabular}           & \begin{tabular}[c]{@{}l@{}}102.36\\ (7.89)\end{tabular}                        \\ \hline
\begin{tabular}[c]{@{}l@{}}Step\\ Length\end{tabular}      & \begin{tabular}[c]{@{}l@{}}69.46\\ (8.01)\end{tabular}                      & \begin{tabular}[c]{@{}l@{}}48.11\\ (6.62)\end{tabular}                     & \begin{tabular}[c]{@{}l@{}}66.02\\ (7.23)\end{tabular}            & \begin{tabular}[c]{@{}l@{}}55.19\\ (9.04)\end{tabular}            & \begin{tabular}[c]{@{}l@{}}53.86\\ (5.94)\end{tabular}                          \\ \hline
\begin{tabular}[c]{@{}l@{}}Stride\\ Length\end{tabular}    & \begin{tabular}[c]{@{}l@{}}78.93\\ (11.19)\end{tabular}                     & \begin{tabular}[c]{@{}l@{}}69.38\\ (8.04)\end{tabular}                    & \begin{tabular}[c]{@{}l@{}}78.11\\ (9.75)\end{tabular}         & \begin{tabular}[c]{@{}l@{}}68.77\\ (9.06)\end{tabular}           & \begin{tabular}[c]{@{}l@{}}70.01\\ (11.23)\end{tabular}                        \\ \hline
\begin{tabular}[c]{@{}l@{}}Step \\ Time\end{tabular}       & \begin{tabular}[c]{@{}l@{}}1.03\\ (0.56)\end{tabular}                        & \begin{tabular}[c]{@{}l@{}}1.07\\ (0.42)\end{tabular}                      & \begin{tabular}[c]{@{}l@{}}0.91\\ (0.51)\end{tabular}            & \begin{tabular}[c]{@{}l@{}}0.99\\ (0.37)\end{tabular}             & \begin{tabular}[c]{@{}l@{}}0.98\\ (0.29)\end{tabular}                           \\ \hline
\begin{tabular}[c]{@{}l@{}}Cycle\\ Time\end{tabular}       & \begin{tabular}[c]{@{}l@{}}1.87\\ (0.94)\end{tabular}                       & \begin{tabular}[c]{@{}l@{}}1.99\\ (0.91)\end{tabular}                      & \begin{tabular}[c]{@{}l@{}}1.81\\ (0.77)\end{tabular}            & \begin{tabular}[c]{@{}l@{}}1.91\\ (0.65)\end{tabular}             & \begin{tabular}[c]{@{}l@{}}1.89\\ (0.82)\end{tabular}                          \\ \hline
\begin{tabular}[c]{@{}l@{}}Swing \\ Time\end{tabular}      & \begin{tabular}[c]{@{}l@{}}0.39\\ (0.07)\end{tabular}                       & \begin{tabular}[c]{@{}l@{}}0.48\\ (0.09)\end{tabular}                       & \begin{tabular}[c]{@{}l@{}}0.36\\ (0.05)\end{tabular}            & \begin{tabular}[c]{@{}l@{}}0.42\\ (0.06)\end{tabular}             & \begin{tabular}[c]{@{}l@{}}0.41\\ (0.05)\end{tabular}                          \\ \hline
\end{tabular}
\end{table}

\subsubsection{Rhythmic Vibrotactile vs. VR Baseline}
We obtained a significant increase in rhythmic (\textit{M} = 113.45, \textit{SD} = 13.68) compared to VR baseline without vibrotactile feedback (\textit{M} = 96.69, \textit{SD} = 13); \textit{t}(17) = 3.81, \textit{p} $<$ .001, \textit{d} = 0.38. We also observed a significant increase (\textit{p} $<$ .001) in cadence, step length, and stride length in rhythmic than VR baseline without vibrotactile feedback condition. However, results indicated a significant reduction (\textit{p} $<$ .001) in step time, cycle time, and swing time in rhythmic compared to VR baseline without vibrotactile feedback condition. Consequently, rhythmic vibrotactile feedback exceeded the VR baseline condition in terms of gait performance.

Fig. 5 depicts the comparisons of walking velocity across five distinct study conditions for the MI group. Table 2 displays the relative mean and standard deviation (SD) for the five research conditions for the remaining gait metrics that showed a significant improvement.

\subsection{Participants Without MI: Within-Group Comparisons}
\subsubsection{Non-VR Baseline vs. VR Baseline}
Walking velocity was significantly lower in VR baseline without vibrotactile feedback (\textit{M} = 128.22, \textit{SD} = 5.58) than the non-VR baseline without vibrotactile feedback condition (\textit{M} = 144.2, \textit{SD} = 8.25); \textit{t}(20) = 10.08, \textit{p} $<$ .001, \textit{d} = 0.14. We also noticed that cadence, step length, and stride length for non-VR baseline were significantly increased (\textit{p} $<$ .001) than VR baseline without vibrotactile feedback condition. However, we observed a significant reduction (\textit{p} $<$ .001) in step time, cycle time, and swing time in non-VR baseline compared to VR baseline without vibrotactile feedback condition. The findings demonstrated that participants without MI experienced gait disturbances in VR.


\begin{table}[ht]
\caption{Gait metrics in five conditions for participants without MI}
    \label{tab:my_label}

\begin{tabular}{|l|l|l|l|l|l|}
\hline
\begin{tabular}[c]{@{}l@{}}Gait \\ Metrics\end{tabular}         & \begin{tabular}[c]{@{}l@{}}Non-VR \\ baseline\\ \\ Mean\\ (SD)\end{tabular} & \begin{tabular}[c]{@{}l@{}}VR\\ baseline\\ \\ Mean\\ (SD)\end{tabular} & \begin{tabular}[c]{@{}l@{}}Spatial\\ \\ Mean\\ (SD)\end{tabular} & \begin{tabular}[c]{@{}l@{}}Rhythmic\\ \\ Mean\\ (SD)\end{tabular} & \begin{tabular}[c]{@{}l@{}}Static \\ \\ Mean\\ (SD)\end{tabular} \\ \hline
Cadence                                                         & \begin{tabular}[c]{@{}l@{}}128.37\\ (7.18)\end{tabular}                     & \begin{tabular}[c]{@{}l@{}}106.63\\ (4.33)\end{tabular}                    & \begin{tabular}[c]{@{}l@{}}126.52\\ (9.47)\end{tabular}         & \begin{tabular}[c]{@{}l@{}}117.89\\ (6.51)\end{tabular}               & \begin{tabular}[c]{@{}l@{}}118.44\\ (12.35)\end{tabular}                         \\ \hline
\begin{tabular}[c]{@{}l@{}}Step\\ Length\end{tabular}    & \begin{tabular}[c]{@{}l@{}}77.28\\ (7.19)\end{tabular}                      & \begin{tabular}[c]{@{}l@{}}54.34\\ (5.09)\end{tabular}                     & \begin{tabular}[c]{@{}l@{}}75.01\\ (7.1)\end{tabular}            & \begin{tabular}[c]{@{}l@{}}66.87\\ (8.12)\end{tabular}                & \begin{tabular}[c]{@{}l@{}}65.42\\ (6.88)\end{tabular}                         \\ \hline

\begin{tabular}[c]{@{}l@{}}Stride\\ Length\end{tabular}  & \begin{tabular}[c]{@{}l@{}}90.78\\ (8.11)\end{tabular}                     & \begin{tabular}[c]{@{}l@{}}71.46\\ (6.89)\end{tabular}                    & \begin{tabular}[c]{@{}l@{}}89.14\\ (7.45)\end{tabular}         & \begin{tabular}[c]{@{}l@{}}81.57\\ (5.78)\end{tabular}               & \begin{tabular}[c]{@{}l@{}}82.84\\ (6.87)\end{tabular}                         \\ \hline

\begin{tabular}[c]{@{}l@{}}Step \\ Time\end{tabular}     & \begin{tabular}[c]{@{}l@{}}1.14\\ (0.09)\end{tabular}                       & \begin{tabular}[c]{@{}l@{}}1.31\\ (0.4)\end{tabular}                      & \begin{tabular}[c]{@{}l@{}}1.11\\ (0.1)\end{tabular}             & \begin{tabular}[c]{@{}l@{}}1.19\\ (0.08)\end{tabular}                 & \begin{tabular}[c]{@{}l@{}}1.18\\ (0.05)\end{tabular}                          \\ \hline
\begin{tabular}[c]{@{}l@{}}Cycle\\ Time\end{tabular}     & \begin{tabular}[c]{@{}l@{}}1.68\\ (0.31)\end{tabular}                       & \begin{tabular}[c]{@{}l@{}}1.99\\ (0.16)\end{tabular}                       & \begin{tabular}[c]{@{}l@{}}1.76\\ (0.19)\end{tabular}            & \begin{tabular}[c]{@{}l@{}}1.88\\ (0.15)\end{tabular}                  & \begin{tabular}[c]{@{}l@{}}1.86\\ (0.27)\end{tabular}                           \\ \hline
\begin{tabular}[c]{@{}l@{}}Swing \\ Time\end{tabular}    & \begin{tabular}[c]{@{}l@{}}0.34\\ (0.09)\end{tabular}                       & \begin{tabular}[c]{@{}l@{}}0.41\\ (0.1)\end{tabular}                      & \begin{tabular}[c]{@{}l@{}}0.33\\ (0.06)\end{tabular}            & \begin{tabular}[c]{@{}l@{}}0.37\\ (0.07)\end{tabular}                 & \begin{tabular}[c]{@{}l@{}}0.36\\ (0.04)\end{tabular}                          \\ \hline
\end{tabular}
\end{table}

\subsubsection{Spatial Vibrotactile vs. VR Baseline}
Experimental results indicated a significant increase in walking velocity in the spatial (\textit{M} = 142.89, \textit{SD} = 11.39) compared to VR baseline without vibrotactile feedback condition (\textit{M} = 128.22, \textit{SD} = 5.58); \textit{t}(20) = 6.87, \textit{p} $<$ .001, \textit{d} = 0.75. Likewise, results also revealed that cadence, step length, and stride length for spatial vibrotactile condition increased significantly (\textit{p} $<$ .001). However, we noticed that step time, cycle time, and swing time were significantly decreased (\textit{p} $<$ .001) in spatial compared to VR baseline without vibrotactile feedback condition. Therefore, spatial vibrotactile feedback produced superior outcomes than the VR baseline condition without vibrotactile feedback.

\subsubsection{Spatial Vibrotactile vs. Static Vibrotactile}
We observed that walking velocity was significantly higher in spatial (\textit{M} = 142.89, \textit{SD} = 11.39) compared to static vibrotactile feedback condition (\textit{M} = 134.62, \textit{SD} = 7.65); \textit{t}(20) = 3.53, \textit{p} $<$ .001, \textit{d} = 0.38. Additionally, we found a  significant increase (\textit{p} $<$ .001) in cadence, step length, and stride length for spatial compared to static vibrotactile feedback condition. However, there was a significant reduction (\textit{p} $<$ .001) in step time, cycle time and swing time in the spatial vibrotactile feedback condition. The results suggested that spatial vibrotactile feedback may be more efficacious than static vibrotactile feedback for gait performance.

\subsubsection{Spatial Vibrotactile vs. Rhythmic Vibrotactile}
Experimental results showed a significant increase in walking velocity in the spatial (\textit{M} = 142.89, \textit{SD} = 11.39) compared to the rhythmic vibrotactile feedback condition (\textit{M} = 134.57, \textit{SD} = 7.19); \textit{t}(20) = 3.36, \textit{p} $<$ .001, \textit{d} = 0.6. Likewise, cadence, step length, and stride length for spatial condition significantly incremented (\textit{p} $<$ .001) than rhythmic vibrotactile feedback condition. However, we observed significant reduction (\textit{p} $<$ .001) in step time, cycle time, and swing time in spatial compared to rhythmic vibrotactile feedback condition. As a result, spatial vibrotactile feedback could be favored more than the rhythmic vibrotactile feedback for gait improvement in VR environments.

\subsubsection{Static Vibrotactile vs. VR Baseline}
Experimental results revealed that walking velocity in static (\textit{M} = 134.62, \textit{SD} = 7.65) was significantly higher compared to VR baseline without vibrotactile feedback condition (\textit{M} = 128.22, \textit{SD} = 5.58); \textit{t}(20) = 2.97, \textit{p} $<$ .001, \textit{d} = 0.3. Likewise, significant increase (\textit{p} $<$ .001) were observed in cadence, step length, and stride length for static vibrotactile feedback than VR baseline without vibrotactile feedback condition. However, we found a significant decline in step time, cycle time, and swing time in static vibrotactile feedback condition (\textit{p} $<$ .001) compared to VR baseline without vibrotactile feedback. Thus, static vibrotactile feedback surpassed the no vibrotactile feedback in VR.

\subsubsection{Static Vibrotactile vs. Rhythmic Vibrotactile}
We found no significant difference in walking velocity between static (\textit{M} = 84.76, \textit{SD} = 13.78) and rhythmic vibrotactile feedback condition (\textit{M} = 82.66, \textit{SD} = 15.69); \textit{t}(20) = 1.11, \textit{p} $=$ .138, \textit{d} = 0.14. Likewise, there was no significant difference for the other gait metrics between static and rhythmic vibrotactile feedback condition. Hence, it was uncertain whether rhythmic or static vibrotactile input would be more effective for gait improvement.

\subsubsection{Rhythmic vibrotactile vs. VR Baseline}
Walking velocity was significantly increased  in rhythmic (\textit{M} = 134.57, \textit{SD} = 7.19) than VR baseline without vibrotactile feedback condition (\textit{M} = 128.22, \textit{SD} = 5.58); \textit{t}(20) = 3.63, \textit{p} $<$ .001, \textit{d} = 0.25. However, step time, cycle time, and swing time significantly decreased (\textit{p} $<$ .001) in rhythmic vibrotactile feedback compared to VR baseline. Also, significant increment (\textit{p} $<$ .001) in cadence, step length, and stride length was observed for rhythmic condition compared to VR baseline without vibrotactile feedback condition. This analysis revealed that rhythmic vibrotactile feedback might be more effective for improving gait in virtual reality environments than the absence of vibrotactile feedback in VR. 

\begin{figure}[ht!]
    \centering
  \includegraphics[width=0.5\textwidth,height=6cm]{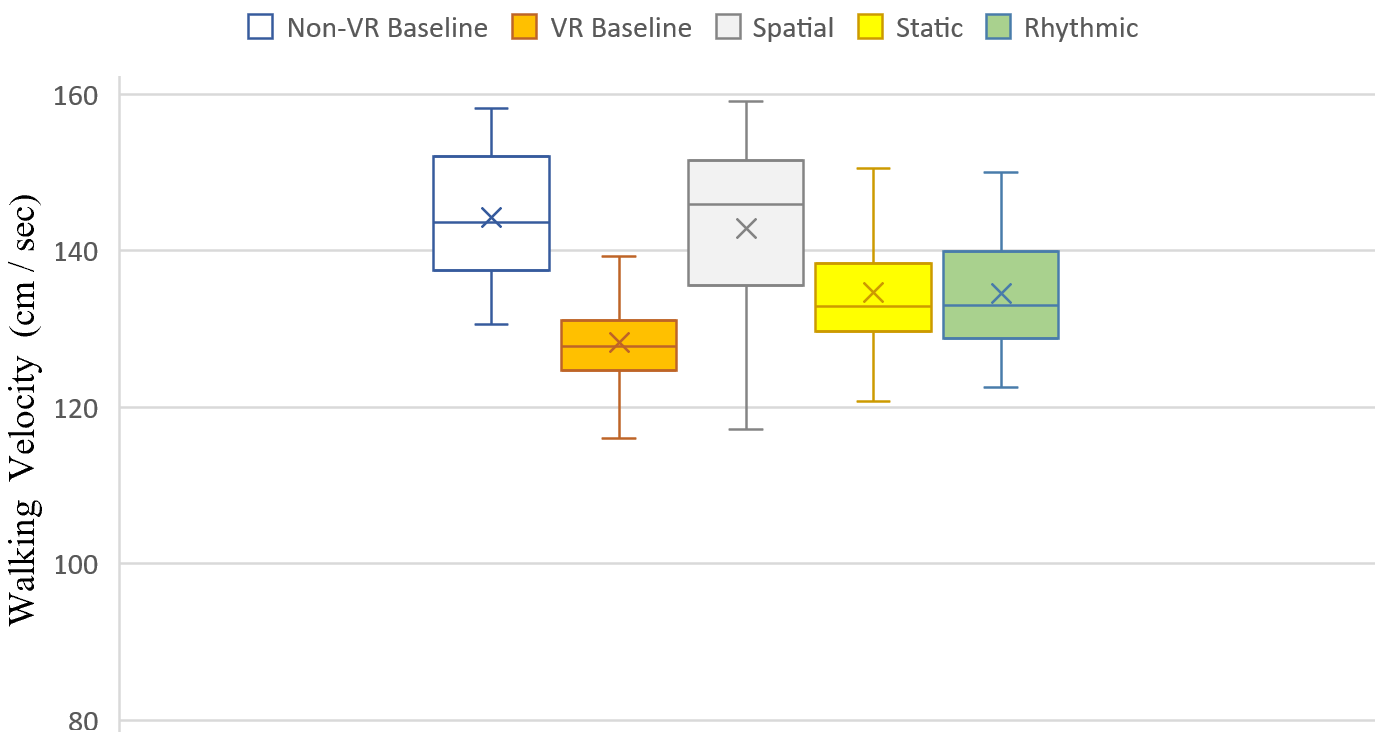}
  \caption{Walking velocity comparison between study conditions for participants without MI.}
\end{figure}


Fig. 6 compares the walking velocity of participants without MI in five different study conditions. Table 3 contains the relative means and standard deviations (SD) for the other six gait metrics that exhibited a statistically significant improvement. Fig. 7 shows comparisons of effect sizes (Cohen's \textit{d}) for walking velocity across various study conditions.

\begin{figure}[ht!]
    \centering
  \includegraphics[width=0.49\textwidth,height=7cm]{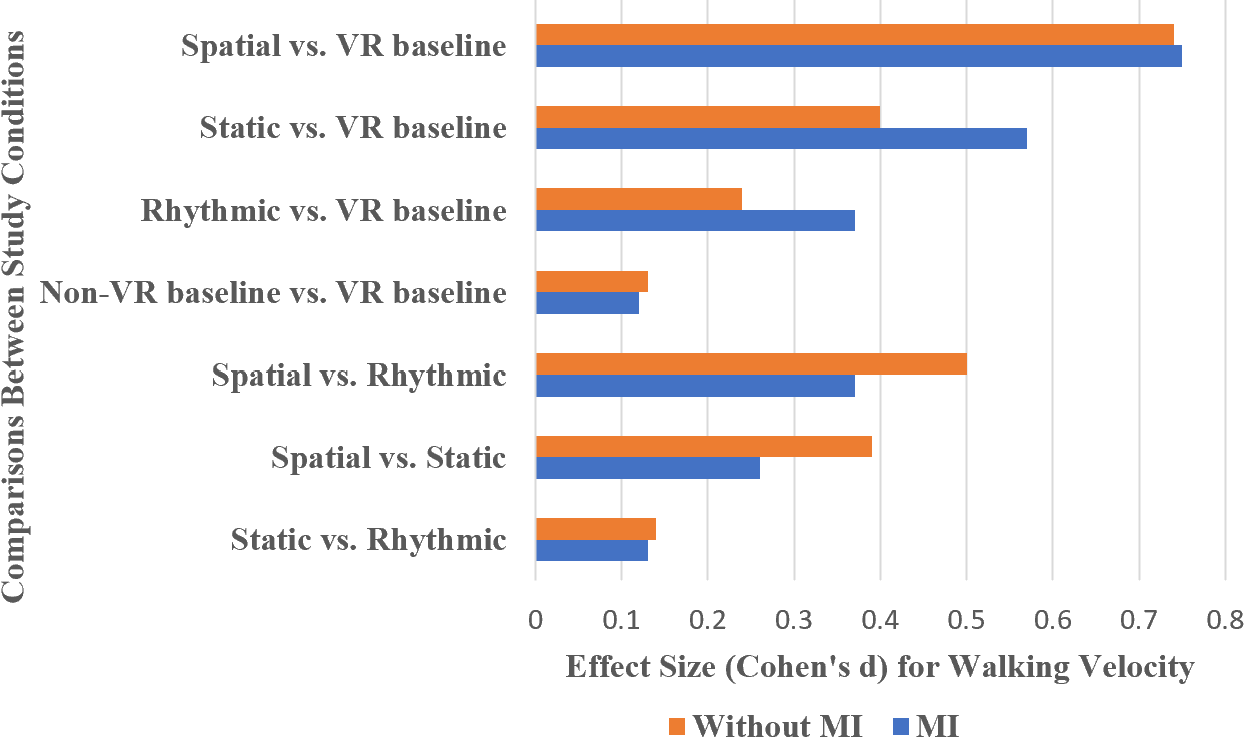}
   \caption{Comparisons of effect size for walking velocity between study conditions for participants with and without MI.}
\end{figure}

\subsection{Between-Group Comparisons}

The mixed-ANOVA and post-hoc two-tailed t-tests revealed that the walking velocity  for participants with MI was significantly lower than that of participants without MI for non-VR baseline condition; \textit{t}(38) = 4.59, \textit{p} $=$ .003; \textit{d} = 1.06, and for VR baseline condition; \textit{t}(38) = 2.88, \textit{p} $=$ .002; \textit{d} = 1.15.
Furthermore, we found that walking velocity was significantly lower for participants with MI than participants without MI for all VR-based vibrotactile feedback conditions: spatial vibrotactile (\textit{t}(38) = 1.21, \textit{p} $=$ .01; \textit{d} = 0.92), static vibrotactile (\textit{t}(38) = 1.93, \textit{p} $=$ .008; \textit{d} = 0.8), and for rhythmic vibrotactile feedback condition (\textit{t}(38) = 1.14, \textit{p} $=$ .02; \textit{d} = 0.77).
For all conditions (non-VR baseline, VR baseline, spatial, static, and rhythmic vibrotactile), participants with MI demonstrated a significant decrease (\textit{p} $<$ .05) in cadence, step length, and stride length compared to participants without MI.

\subsection{Activities-specific Balance Confidence (ABC) Scale}
Two-tailed t-test was performed on the response scores from the Activities-specific Balance Confidence (ABC) Scale between participants with MI ($M$ = 72.64, $SD$ = 19.33) and participants without MI ($M$ = 93.35, $SD$ = 8.06), \textit{t}(38) = 6.19, \textit{p} $<$ .001, \textit{d} = 1.01. The computed mean score on the ABC Scale for participants with MI was 72.64\%, indicating a moderate degree of functioning. However, the computed mean ABC score for those without MI was 91.76\% which demonstrated a high level of functioning. These ratings indicated a significant difference between those with and without MI in terms of physical functioning. 

\subsection{Simulator Sickness Questionnaire}
For all participants, we performed a two-tailed t-test between pre-study SSQ scores and post-study SSQ scores. We did not notice a statistically significant rise in SSQ scores among participants with and without MI. We found \textit{t}(17) = 1.71, \textit{p} = .07, \textit{d} = 0.2 for participants with MI, while \textit{t}(20) = 1.58, \textit{p} = .09, \textit{d} = 0.1 for participants without MI.


\section{DISCUSSION}
\label{sec:discussion}

\subsection{Gait Disturbances in VR Without Vibrotactile Feedback}
Mixed ANOVA and post-hoc t-tests for both participant groups revealed that walking velocity, step length, stride length, cadence, step time, cycle time, and swing time were significantly changed (\textit{p} $<$ .001) in the VR baseline without vibrotactile condition relative to the non-VR baseline without vibrotactile condition. Therefore, we observed that gait disturbances occurred for all individuals in the absence of vibrotactile feedback, supporting our hypothesis H1. Previous research has shown that VR may produce postural instability, which may result in gait disturbances \cite{hollman2007does,riem2020effect,sondell2005altered}.

\subsection{Gait Improvement in VR-based Vibrotactile Feedback Conditions}
For both participants with and without MI, results revealed that all VR-based vibrotactile conditions increased gait performance significantly (\textit{p} $<$ .001) compared to VR baseline condition, which supported our hypothesis H2. In addition, effect size (Cohen's \textit{d} = 0.5) demonstrated that Spatial vibrotactile feedback had a medium effect on both participants with and without MI. The static vibrotactile had a medium effect (Cohen's \textit{d} = 0.5) on participants with MI but a small effect (Cohen's \textit{d} = 0.2) on those without MI. Rhythmic audio had a small effect (Cohen's \textit{d} = 0.2) on both participant groups.  

Spatial vibrotactile feedback performed significantly better (\textit{p} $<$ .001) than other VR-based vibrotactile feedback conditions, supporting our hypothesis H3. Additionally, spatial vibrotactile feedback exhibited a larger effect size than other conditions (Fig. 7). In prior research, simulated spatial vibrotactile feedback was shown to be useful in the real world for improving gait and postural stability due to its superior quality \cite{ballardini2020vibrotactile}. However, the majority of prior research was conducted in non-VR environments, while we evaluated the influence of vibrotactile feedback in VR. Also, earlier studies only examined spatial vibrotactile or only a specific form of vibrotactile feedback, while this research compared three distinct types of vibrotactile feedback in VR.

\subsection{Gait Similarities and Dissimilarities Between Participants With and Without MI}
All five research conditions revealed significant variations in walking velocity, cadence, step length, and stride length between people with and without MI. Other gait metrics, however, did not change significantly between people with and without MI. Thus, our hypothesis H4 was supported by the fact that some gait metrics (e.g., velocity, cadence, step length, stride length) were altered differently between individuals with and without MI, while some other gait metrics were influenced similarly for both groups of participants. These results were somewhat consistent with a prior study by Guo et al. \cite{guo2015mobility} where they also compared the gait metrics of people with and without MI in a VE. They reported walking velocity, step length, and stride length were significantly different between people with and without MI, although there were no significant variations in other gait metrics between the two groups. However, they did not investigate the effect of vibrotactile feedback.

To determine which group had greater gait improvement in response to the vibrotactile feedback conditions, we first subtracted baseline data from each condition. Then, ANOVA and post hoc two-tailed t-tests indicated that the gait improvement of individuals with MI was substantially greater (\textit{p} $<$ .001) than that of those without MI. Impact size, as measured by Cohen's \textit{d} = 0.9, also revealed that people with MI had a greater effect. We anticipated that since people with MI had less gait functioning than those without MI, they could be more amenable to improvement.

\subsection{Cybersickness}
Previous studies found that VR users exposed to virtual worlds for more than 10 minutes might suffer the development of cybersickness \cite{chang2020virtual,kim2021clinical}. Participants in our research wore the HMD for around 45 minutes under four conditions, which raised their likelihood of getting cybersickness symptoms. However, we built the virtual environment without the illusion of self-motion to reduce the likelihood of participants experiencing cybersickness \cite{mccauley1992cybersickness}, and there were resting periods between conditions. Therefore, there was no statistically significant difference between pre-SSQ and post-SSQ ratings. Although we observed that many individuals suffered minor cybersickness after the trials, this did not seem to significantly affect gait.

\section{Limitations}
The participants were informed of the whole research protocol at the outset of the trials. Then we conducted a few trials with them until they were used to the experimental procedure. Before beginning the vibrotactile feedback in VR, we also conducted three baseline trials. However, the baseline might be expanded to more trials. We did not do this since the trials took around an hour to complete and included individuals with MI due to MS who had reduced physical functioning. However, we counterbalanced the feedback conditions in VR to decrease the learning effects.

In our research, the feedback conditions were administered in a counterbalanced sequence, which reduced carryover or practice effects (learning due to repetition) \cite{WinNT5, WinNT4, WinNT6}. Counterbalancing order was also found to be successful in a number of previous studies \cite{millner2020suicide,plechata2019age,sheskin2018thechildlab}. Alternately, we may have administered the vibrotactile feedback conditions in a random sequence to minimize bias. Nevertheless, our research included MS patients who are very prone to fatigue and cybersickness. Consequently, we were particularly worried about the impact of carryover fatigue and cybersickness on performance. Since counterbalancing significantly minimizes fatigue and cybersickness effects \cite{WinNT6}, we preferred counterbalancing over randomization.

Participants wore harnesses throughout the research to prevent falls, which may have improved their gait performance somewhat. To ensure the consistency and safety of the research protocol, we asked all participants to wear harnesses for all baseline and VR trials, regardless of whether they had the risk of falling or not. Consequently, investigations examining gait without a harness may provide slightly different findings.

The duration of the research was long, and each participant had to complete 15 time-consuming GAITRite trials. We discovered that the duration and physical demands of the research might sometimes induce tiredness in participants. To reduce tiredness in our research, individuals were permitted to remove their HMD and relax between trials and conditions. This respite and removal of the HMD may have enabled them to reestablish spatial orientation inside the room, resulting in somewhat biased data.

The vibrating motors used for vibrotactile feedback created noise. So it's hard to say if the sounds it made contributed to the gait improvement, if it was the actual vibration, or both. More research will be needed to verify this.

The "rhythmic" vibrotactile feedback was provided at one-second intervals. We did not examine this feedback condition for different time periods (e.g., two-second). Therefore, studies that provide "rhythmic" vibrotactile input with varying time intervals may produce slightly different findings for this particular circumstance.

For the static vibrotactile feedback, participants may have experienced fatigue as the vibration was continually played. However, the fatigue impact for this circumstance was not measured.

In our research, the non-VR baseline was always performed first, which may have had an effect on the walking speed for this condition. However, we wanted to limit the learning impact of VR circumstances by introducing sufficient baseline tasks prior to VR conditions.

In our study, more females than males participated in the MI group participants. This is due to the fact that we recruited from the MS community, which is statistically more prevalent in females \cite{WinNT}. Numerous prior studies have shown no significant gender impact on balance \cite{kahraman2018gender, faraldo2012influence,schedler2019age}. In our future studies, we want to examine the gender influence on gait metrics in VR.

During the VR intervention, we assessed gait performance. We did not assess post-study gait effects. Our objective was accessibility rather than rehabilitation. Therefore we only examined gait outside of VR as a baseline and during VR immersion.

There were five distinct research conditions. We conducted three trials for each research condition, resulting in a total of 15 trials per participant, and we gathered different data files for each trial. We conducted a total of 585 trials with our 39 participants. The HMD display failed during three trials involving three individuals (MI group:2, without MI group: 1). Restarting the "Vive wireless app" always resolved the problem. We repeated the three trials and omitted the three flawed data files.

Due to COVID-19 and our intended test group, which included individuals with MI caused by MS, the recruiting procedure was challenging. Because many prospective volunteers had impaired immune systems, which placed them at significant risk for COVID-19. Consequently, they excluded themselves from the research. We could have recruited more participants if the research had been conducted outside of COVID-19.

\section{Conclusion and future work}
\label{sec:conclusion}

In this study, we assessed the effects of several vibrotactile feedback modalities (spatial, static, and rhythmic) on gait in VR. In our research, all vibrotactile feedback conditions substantially improved gait in VR. Spatial vibrotactile feedback outperformed rhythmic and static vibrotactile significantly. There was no statistically significant difference between rhythmic and static vibrotactile feedback. Researchers will be better able to comprehend the various types of vibrotactile input for improving gait in an HMD-based VE as a consequence of these findings. In addition, this study may assist developers in creating VR experiences that are more accessible and useful for those with and without mobility issues. Future research will include evaluating the efficacy of other feedback modalities to make real walking in VR more accessible.

\ifCLASSOPTIONcompsoc
  \section*{Acknowledgments}
\else
  \section*{Acknowledgment}
\fi

The authors would like to express their gratitude to everyone who participated in this study. This project was funded by a grant from the National Science Foundation (IIS 2007041).

\ifCLASSOPTIONcaptionsoff
  \newpage
\fi



\bibliographystyle{IEEEtran}
\bibliography{IEEEabrv,template}

\begin{thebibliography}{10}
\providecommand{\url}[1]{#1}
\csname url@samestyle\endcsname
\providecommand{\newblock}{\relax}
\providecommand{\bibinfo}[2]{#2}
\providecommand{\BIBentrySTDinterwordspacing}{\spaceskip=0pt\relax}
\providecommand{\BIBentryALTinterwordstretchfactor}{4}
\providecommand{\BIBentryALTinterwordspacing}{\spaceskip=\fontdimen2\font plus
\BIBentryALTinterwordstretchfactor\fontdimen3\font minus
  \fontdimen4\font\relax}
\providecommand{\BIBforeignlanguage}[2]{{%
\expandafter\ifx\csname l@#1\endcsname\relax
\typeout{** WARNING: IEEEtran.bst: No hyphenation pattern has been}%
\typeout{** loaded for the language `#1'. Using the pattern for}%
\typeout{** the default language instead.}%
\else
\language=\csname l@#1\endcsname
\fi
#2}}
\providecommand{\BIBdecl}{\relax}
\BIBdecl

\bibitem{agrawal2009disorders}
Y.~Agrawal, J.~P. Carey, C.~C. Della~Santina, M.~C. Schubert, and L.~B. Minor,
  ``Disorders of balance and vestibular function in us adults: data from the
  national health and nutrition examination survey, 2001-2004,'' \emph{Archives
  of internal medicine}, vol. 169, no.~10, pp. 938--944, 2009.

\bibitem{ferdous2016visual}
S.~M.~S. Ferdous, I.~M. Arafat, and J.~Quarles, ``Visual feedback to improve
  the accessibility of head-mounted displays for persons with balance
  impairments,'' in \emph{2016 IEEE Symposium on 3D User Interfaces
  (3DUI)}.\hskip 1em plus 0.5em minus 0.4em\relax IEEE, 2016, pp. 121--128.

\bibitem{ferdous2018investigating}
S.~M.~S. Ferdous, T.~I. Chowdhury, I.~M. Arafat, and J.~Quarles,
  ``Investigating the reason for increased postural instability in virtual
  reality for persons with balance impairments,'' in \emph{Proceedings of the
  24th ACM Symposium on Virtual Reality Software and Technology}, 2018, pp.
  1--7.

\bibitem{guo2013effects}
R.~Guo, G.~Samaraweera, and J.~Quarles, ``The effects of ves on mobility
  impaired users: Presence, gait, and physiological response,'' in
  \emph{Proceedings of the 19th ACM Symposium on Virtual Reality Software and
  Technology}, 2013, pp. 59--68.

\bibitem{samaraweera2013latency}
G.~Samaraweera, R.~Guo, and J.~Quarles, ``Latency and avatars in virtual
  environments and the effects on gait for persons with mobility impairments,''
  in \emph{2013 IEEE Symposium on 3D User Interfaces (3DUI)}.\hskip 1em plus
  0.5em minus 0.4em\relax IEEE, 2013, pp. 23--30.

\bibitem{WinNT}
``{[n. d.]} people with disabilities in the world,''
  \url{https://www.un.org/development/desa/disabilities/resources/factsheet-on-persons-with-disabilities.html},
  accessed: 2021-08-30.

\bibitem{franco2012ibalance}
C.~Franco, A.~Fleury, P.-Y. Gum{\'e}ry, B.~Diot, J.~Demongeot, and
  N.~Vuillerme, ``ibalance-abf: a smartphone-based audio-biofeedback balance
  system,'' \emph{IEEE transactions on biomedical engineering}, vol.~60, no.~1,
  pp. 211--215, 2012.

\bibitem{sienko2017role}
K.~Sienko, S.~Whitney, W.~Carender, and C.~Wall~III, ``The role of sensory
  augmentation for people with vestibular deficits: real-time balance aid
  and/or rehabilitation device?'' \emph{Journal of Vestibular Research},
  vol.~27, no.~1, pp. 63--76, 2017.

\bibitem{velazquez2010wearable}
R.~Vel{\'a}zquez, ``Wearable assistive devices for the blind,'' in
  \emph{Wearable and autonomous biomedical devices and systems for smart
  environment}.\hskip 1em plus 0.5em minus 0.4em\relax Springer, 2010, pp.
  331--349.

\bibitem{thikey2011need}
H.~Thikey, F.~van Wjick, M.~Grealy, and P.~Rowe, ``A need for meaningful visual
  feedback of lower extremity function after stroke,'' in \emph{2011 5th
  International Conference on Pervasive Computing Technologies for Healthcare
  (PervasiveHealth) and Workshops}.\hskip 1em plus 0.5em minus 0.4em\relax
  IEEE, 2011, pp. 379--383.

\bibitem{vcakrt2010exercise}
O.~{\v{C}}akrt, M.~Chovanec, T.~Funda, P.~Kalitov{\'a}, J.~Betka,
  E.~Zv{\v{e}}{\v{r}}ina, P.~Kol{\'a}{\v{r}}, and J.~Je{\v{r}}{\'a}bek,
  ``Exercise with visual feedback improves postural stability after vestibular
  schwannoma surgery,'' \emph{European archives of oto-rhino-laryngology}, vol.
  267, no.~9, pp. 1355--1360, 2010.

\bibitem{sutbeyaz2007mirror}
S.~S{\"u}tbeyaz, G.~Yavuzer, N.~Sezer, and B.~F. Koseoglu, ``Mirror therapy
  enhances lower-extremity motor recovery and motor functioning after stroke: a
  randomized controlled trial,'' \emph{Archives of physical medicine and
  rehabilitation}, vol.~88, no.~5, pp. 555--559, 2007.

\bibitem{gibbs2022comparison}
J.~K. Gibbs, M.~Gillies, and X.~Pan, ``A comparison of the effects of haptic
  and visual feedback on presence in virtual reality,'' \emph{International
  Journal of Human-Computer Studies}, vol. 157, p. 102717, 2022.

\bibitem{soltani2020influence}
P.~Soltani and R.~Andrade, ``The influence of virtual reality head-mounted
  displays on balance outcomes and training paradigms: A systematic review.''
  \emph{Frontiers in sports and active living}, vol.~2, p. 233, 2020.

\bibitem{martinez2018analysing}
A.~Martinez, A.~I. Paganelli, and A.~Raposo, ``Analysing balance loss in vr
  interaction with hmds,'' \emph{Journal on Interactive Systems}, vol.~9,
  no.~2, 2018.

\bibitem{murata2004effects}
A.~Murata, ``Effects of duration of immersion in a virtual reality environment
  on postural stability,'' \emph{International Journal of Human-Computer
  Interaction}, vol.~17, no.~4, pp. 463--477, 2004.

\bibitem{hollman2007does}
J.~H. Hollman, R.~H. Brey, T.~J. Bang, and K.~R. Kaufman, ``Does walking in a
  virtual environment induce unstable gait?: An examination of vertical ground
  reaction forces,'' \emph{Gait \& posture}, vol.~26, no.~2, pp. 289--294,
  2007.

\bibitem{riem2020effect}
L.~I. Riem, B.~D. Schmit, and S.~A. Beardsley, ``The effect of discrete visual
  perturbations on balance control during gait,'' in \emph{2020 42nd Annual
  International Conference of the IEEE Engineering in Medicine \& Biology
  Society (EMBC)}.\hskip 1em plus 0.5em minus 0.4em\relax IEEE, 2020, pp.
  3162--3165.

\bibitem{sondell2005altered}
B.~Sondell, L.~Nyberg, S.~Eriksson, B.~Engstr{\"o}m, A.~Backman, K.~Holmlund,
  G.~Bucht, and L.~Lundin-Olsson, ``Altered walking pattern in a virtual
  environment,'' \emph{Presence}, vol.~14, no.~2, pp. 191--197, 2005.

\bibitem{horsak2021overground}
B.~Horsak, M.~Simonlehner, L.~Sch{\"o}ffer, B.~Dumphart, A.~Jalaeefar, and
  M.~Husinsky, ``Overground walking in a fully immersive virtual reality: A
  comprehensive study on the effects on full-body walking biomechanics,''
  \emph{Frontiers in bioengineering and biotechnology}, vol.~9, 2021.

\bibitem{canessa2019comparing}
A.~Canessa, P.~Casu, F.~Solari, and M.~Chessa, ``Comparing real walking in
  immersive virtual reality and in physical world using gait analysis.'' in
  \emph{VISIGRAPP (2: HUCAPP)}, 2019, pp. 121--128.

\bibitem{martelli2019gait}
D.~Martelli, B.~Xia, A.~Prado, and S.~K. Agrawal, ``Gait adaptations during
  overground walking and multidirectional oscillations of the visual field in a
  virtual reality headset,'' \emph{Gait \& posture}, vol.~67, pp. 251--256,
  2019.

\bibitem{de2016effect}
I.~J. De~Rooij, I.~G. Van De~Port, and J.-W.~G. Meijer, ``Effect of virtual
  reality training on balance and gait ability in patients with stroke:
  systematic review and meta-analysis,'' \emph{Physical therapy}, vol.~96,
  no.~12, pp. 1905--1918, 2016.

\bibitem{meldrum2012effectiveness}
D.~Meldrum, S.~Herdman, R.~Moloney, D.~Murray, D.~Duffy, K.~Malone, H.~French,
  S.~Hone, R.~Conroy, and R.~McConn-Walsh, ``Effectiveness of conventional
  versus virtual reality based vestibular rehabilitation in the treatment of
  dizziness, gait and balance impairment in adults with unilateral peripheral
  vestibular loss: a randomised controlled trial,'' \emph{BMC Ear, Nose and
  Throat Disorders}, vol.~12, no.~1, pp. 1--8, 2012.

\bibitem{park2015effects}
E.-C. Park, S.-G. Kim, and C.-W. Lee, ``The effects of virtual reality game
  exercise on balance and gait of the elderly,'' \emph{Journal of physical
  therapy science}, vol.~27, no.~4, pp. 1157--1159, 2015.

\bibitem{cho2016treadmill}
C.~Cho, W.~Hwang, S.~Hwang, and Y.~Chung, ``Treadmill training with virtual
  reality improves gait, balance, and muscle strength in children with cerebral
  palsy,'' \emph{The Tohoku journal of experimental medicine}, vol. 238, no.~3,
  pp. 213--218, 2016.

\bibitem{duque2013effects}
G.~Duque, D.~Boersma, G.~Loza-Diaz, S.~Hassan, H.~Suarez, D.~Geisinger,
  P.~Suriyaarachchi, A.~Sharma, and O.~Demontiero, ``Effects of balance
  training using a virtual-reality system in older fallers,'' \emph{Clinical
  interventions in aging}, vol.~8, p. 257, 2013.

\bibitem{bergeron2015use}
M.~Bergeron, C.~L. Lortie, and M.~J. Guitton, ``Use of virtual reality tools
  for vestibular disorders rehabilitation: a comprehensive analysis,''
  \emph{Advances in medicine}, vol. 2015, 2015.

\bibitem{walker2010virtual}
M.~L. Walker, S.~I. Ringleb, G.~C. Maihafer, R.~Walker, J.~R. Crouch,
  B.~Van~Lunen, and S.~Morrison, ``Virtual reality--enhanced partial body
  weight--supported treadmill training poststroke: feasibility and
  effectiveness in 6 subjects,'' \emph{Archives of physical medicine and
  rehabilitation}, vol.~91, no.~1, pp. 115--122, 2010.

\bibitem{janeh2019gait}
O.~Janeh, O.~Fr{\"u}ndt, B.~Sch{\"o}nwald, A.~Gulberti, C.~Buhmann, C.~Gerloff,
  F.~Steinicke, and M.~P{\"o}tter-Nerger, ``Gait training in virtual reality:
  short-term effects of different virtual manipulation techniques in
  parkinson’s disease,'' \emph{Cells}, vol.~8, no.~5, p. 419, 2019.

\bibitem{janeh2021review}
O.~Janeh and F.~Steinicke, ``A review of the potential of virtual walking
  techniques for gait rehabilitation,'' \emph{Frontiers in Human Neuroscience},
  vol.~15, 2021.

\bibitem{winter2021immersive}
C.~Winter, F.~Kern, D.~Gall, M.~E. Latoschik, P.~Pauli, and I.~K{\"a}thner,
  ``Immersive virtual reality during gait rehabilitation increases walking
  speed and motivation: a usability evaluation with healthy participants and
  patients with multiple sclerosis and stroke,'' \emph{Journal of
  neuroengineering and rehabilitation}, vol.~18, no.~1, pp. 1--14, 2021.

\bibitem{guo2015mobility}
R.~Guo, G.~Samaraweera, and J.~Quarles, ``Mobility impaired users respond
  differently than healthy users in virtual environments,'' \emph{Computer
  Animation and Virtual Worlds}, vol.~26, no.~5, pp. 509--526, 2015.

\bibitem{lott2003effect}
A.~Lott, E.~Bisson, Y.~Lajoie, J.~McComas, and H.~Sveistrup, ``The effect of
  two types of virtual reality on voluntary center of pressure displacement,''
  \emph{Cyberpsychology \& behavior}, vol.~6, no.~5, pp. 477--485, 2003.

\bibitem{epure2014effect}
P.~Epure, C.~Gheorghe, T.~Nissen, L.-O. Toader, A.~Nicolae, S.~S. Nielsen,
  D.~J.~R. Christensen, A.~L. Brooks, and E.~Petersson, ``Effect of the oculus
  rift head mounted display on postural stability,'' in \emph{The 10th
  International Conference on Disability Virtual Reality \& Associated
  Technologies: Proceedings}.\hskip 1em plus 0.5em minus 0.4em\relax Reading
  University Press, 2014, pp. 119--127.

\bibitem{robert2016effect}
M.~T. Robert, L.~Ballaz, and M.~Lemay, ``The effect of viewing a virtual
  environment through a head-mounted display on balance,'' \emph{Gait \&
  posture}, vol.~48, pp. 261--266, 2016.

\bibitem{horlings2009influence}
C.~G. Horlings, M.~G. Carpenter, U.~M. K{\"u}ng, F.~Honegger, B.~Wiederhold,
  and J.~H. Allum, ``Influence of virtual reality on postural stability during
  movements of quiet stance,'' \emph{Neuroscience letters}, vol. 451, no.~3,
  pp. 227--231, 2009.

\bibitem{samaraweera2015applying}
G.~Samaraweera, A.~Perdomo, and J.~Quarles, ``Applying latency to half of a
  self-avatar's body to change real walking patterns,'' in \emph{2015 IEEE
  Virtual Reality (VR)}.\hskip 1em plus 0.5em minus 0.4em\relax IEEE, 2015, pp.
  89--96.

\bibitem{alahakone2010real}
A.~U. Alahakone and S.~A. Senanayake, ``A real-time system with assistive
  feedback for postural control in rehabilitation,'' \emph{IEEE/ASME
  Transactions on Mechatronics}, vol.~15, no.~2, pp. 226--233, 2010.

\bibitem{chiari2005audio}
L.~Chiari, M.~Dozza, A.~Cappello, F.~B. Horak, V.~Macellari, and D.~Giansanti,
  ``Audio-biofeedback for balance improvement: an accelerometry-based system,''
  \emph{IEEE transactions on biomedical engineering}, vol.~52, no.~12, pp.
  2108--2111, 2005.

\bibitem{sienko2012biofeedback}
K.~H. Sienko, M.~D. Balkwill, and C.~Wall, ``Biofeedback improves postural
  control recovery from multi-axis discrete perturbations,'' \emph{Journal of
  neuroengineering and rehabilitation}, vol.~9, no.~1, pp. 1--11, 2012.

\bibitem{henry2019age}
M.~Henry and S.~Baudry, ``Age-related changes in leg proprioception:
  implications for postural control,'' \emph{Journal of neurophysiology}, vol.
  122, no.~2, pp. 525--538, 2019.

\bibitem{wannstedt1978use}
G.~T. Wannstedt and R.~M. Herman, ``Use of augmented sensory feedback to
  achieve symmetrical standing,'' \emph{Physical Therapy}, vol.~58, no.~5, pp.
  553--559, 1978.

\bibitem{sienko2018potential}
K.~H. Sienko, R.~D. Seidler, W.~J. Carender, A.~D. Goodworth, S.~L. Whitney,
  and R.~J. Peterka, ``Potential mechanisms of sensory augmentation systems on
  human balance control,'' \emph{Frontiers in neurology}, vol.~9, p. 944, 2018.

\bibitem{goodworth2009influence}
A.~D. Goodworth, C.~Wall~III, and R.~J. Peterka, ``Influence of feedback
  parameters on performance of a vibrotactile balance prosthesis,'' \emph{IEEE
  Transactions on Neural Systems and Rehabilitation Engineering}, vol.~17,
  no.~4, pp. 397--408, 2009.

\bibitem{wall2009vibrotactile}
C.~Wall~III, D.~M. Wrisley, and K.~D. Statler, ``Vibrotactile tilt feedback
  improves dynamic gait index: a fall risk indicator in older adults,''
  \emph{Gait \& posture}, vol.~30, no.~1, pp. 16--21, 2009.

\bibitem{rust2020benefits}
H.~Rust, N.~Lutz, V.~Zumbrunnen, M.~Imhof, {\"O}.~Yaldizli, V.~Haller, and
  J.~H. Allum, ``Benefits of short-term training with vibrotactile biofeedback
  of trunk sway on balance control in multiple sclerosis,'' \emph{Physical
  Medicine and Rehabilitation Research}, vol.~5, no.~1, pp. 1--10, 2020.

\bibitem{ballardini2020vibrotactile}
G.~Ballardini, V.~Florio, A.~Canessa, G.~Carlini, P.~Morasso, and M.~Casadio,
  ``Vibrotactile feedback for improving standing balance,'' \emph{Frontiers in
  bioengineering and biotechnology}, vol.~8, p.~94, 2020.

\bibitem{kingma2019vibrotactile}
H.~Kingma, L.~Felipe, M.-C. Gerards, P.~Gerits, N.~Guinand, A.~Perez-Fornos,
  V.~Demkin, and R.~Van De~Berg, ``Vibrotactile feedback improves balance and
  mobility in patients with severe bilateral vestibular loss,'' \emph{Journal
  of neurology}, vol. 266, no.~1, pp. 19--26, 2019.

\bibitem{mahmud2022auditory}
M.~R. Mahmud, M.~Stewart, A.~Cordova, and J.~Quarles, ``Auditory feedback for
  standing balance improvement in virtual reality,'' in \emph{2022 IEEE
  Conference on Virtual Reality and 3D User Interfaces (VR)}.\hskip 1em plus
  0.5em minus 0.4em\relax IEEE, 2022, pp. 782--791.

\bibitem{stevens2016auditory}
M.~N. Stevens, D.~L. Barbour, M.~P. Gronski, and T.~E. Hullar, ``Auditory
  contributions to maintaining balance,'' \emph{Journal of Vestibular
  Research}, vol.~26, no. 5-6, pp. 433--438, 2016.

\bibitem{gandemer2017spatial}
L.~Gandemer, G.~Parseihian, R.~Kronland-Martinet, and C.~Bourdin, ``Spatial
  cues provided by sound improve postural stabilization: evidence of a spatial
  auditory map?'' \emph{Frontiers in neuroscience}, vol.~11, p. 357, 2017.

\bibitem{ross2016auditory}
J.~Ross, O.~Will, Z.~McGann, and R.~Balasubramaniam, ``Auditory white noise
  reduces age-related fluctuations in balance,'' \emph{Neuroscience letters},
  vol. 630, pp. 216--221, 2016.

\bibitem{cornwell2020walking}
T.~Cornwell, J.~Woodward, M.~Wu, B.~Jackson, P.~Souza, J.~Siegel, S.~Dhar,
  K.~E. Gordon \emph{et~al.}, ``Walking with ears: altered auditory feedback
  impacts gait step length in older adults,'' \emph{Frontiers in Sports and
  Active Living}, vol.~2, p.~38, 2020.

\bibitem{ghai2018effect}
S.~Ghai, I.~Ghai, and A.~O. Effenberg, ``Effect of rhythmic auditory cueing on
  aging gait: a systematic review and meta-analysis,'' \emph{Aging and
  disease}, vol.~9, no.~5, p. 901, 2018.

\bibitem{helps2014different}
S.~K. Helps, S.~Bamford, E.~J. Sonuga-Barke, and G.~B. S{\"o}derlund,
  ``Different effects of adding white noise on cognitive performance of sub-,
  normal and super-attentive school children,'' \emph{PloS one}, vol.~9,
  no.~11, p. e112768, 2014.

\bibitem{chong2020audio}
U.~Chong and S.~Alimardanov, ``Audio augmented reality using unity for marine
  tourism,'' in \emph{International Conference on Intelligent Human Computer
  Interaction}.\hskip 1em plus 0.5em minus 0.4em\relax Springer, 2020, pp.
  303--311.

\bibitem{pinkl2020spatialized}
J.~Pinkl and M.~Cohen, ``Spatialized ar polyrhythmic metronome using bose
  frames eyewear,'' 2020.

\bibitem{kim2019immersive}
H.~Kim, L.~Remaggi, P.~J. Jackson, and A.~Hilton, ``Immersive spatial audio
  reproduction for vr/ar using room acoustic modelling from 360 images,'' in
  \emph{2019 IEEE Conference on Virtual Reality and 3D User Interfaces
  (VR)}.\hskip 1em plus 0.5em minus 0.4em\relax IEEE, 2019, pp. 120--126.

\bibitem{mccormack2022parametric}
L.~McCormack, A.~Politis, R.~Gonzalez, T.~Lokki, and V.~Pulkki, ``Parametric
  ambisonic encoding of arbitrary microphone arrays,'' \emph{IEEE/ACM
  Transactions on Audio, Speech, and Language Processing}, 2022.

\bibitem{powell1995activities}
L.~E. Powell and A.~M. Myers, ``The activities-specific balance confidence
  (abc) scale,'' \emph{The Journals of Gerontology Series A: Biological
  Sciences and Medical Sciences}, vol.~50, no.~1, pp. M28--M34, 1995.

\bibitem{kennedy1993simulator}
R.~S. Kennedy, N.~E. Lane, K.~S. Berbaum, and M.~G. Lilienthal, ``Simulator
  sickness questionnaire: An enhanced method for quantifying simulator
  sickness,'' \emph{The international journal of aviation psychology}, vol.~3,
  no.~3, pp. 203--220, 1993.

\bibitem{steffen2002age}
T.~M. Steffen, T.~A. Hacker, and L.~Mollinger, ``Age-and gender-related test
  performance in community-dwelling elderly people: Six-minute walk test, berg
  balance scale, timed up \& go test, and gait speeds,'' \emph{Physical
  therapy}, vol.~82, no.~2, pp. 128--137, 2002.

\bibitem{WinNT3}
``{[GaitRite Manual]} details gait parameters for gaitrite walkway system.''
  \url{https://www.procarebv.nl/wp-content/uploads/2017/01/Technische-aspecten-GAITrite-Walkway-System.pdf},
  accessed: 2021-08-30.

\bibitem{chang2020virtual}
E.~Chang, H.~T. Kim, and B.~Yoo, ``Virtual reality sickness: a review of causes
  and measurements,'' \emph{International Journal of Human--Computer
  Interaction}, vol.~36, no.~17, pp. 1658--1682, 2020.

\bibitem{kim2021clinical}
H.~Kim, D.~J. Kim, W.~H. Chung, K.-A. Park, J.~D. Kim, D.~Kim, K.~Kim, and
  H.~J. Jeon, ``Clinical predictors of cybersickness in virtual reality (vr)
  among highly stressed people,'' \emph{Scientific reports}, vol.~11, no.~1,
  pp. 1--11, 2021.

\bibitem{mccauley1992cybersickness}
M.~E. McCauley and T.~J. Sharkey, ``Cybersickness: Perception of self-motion in
  virtual environments,'' \emph{Presence: Teleoperators \& Virtual
  Environments}, vol.~1, no.~3, pp. 311--318, 1992.

\bibitem{WinNT5}
``{[Counterbalancing]} one of the best ways to avoid pitfalls of repeated
  measure designs.''
  \url{https://explorable.com/counterbalanced-measures-design#:~:text=Experiments%20conducted%20with%20a%20counterbalanced%20measures%20design%20are,subjects%20are%20exposed%20to%20all%20of%20the%20treatments.},
  accessed: 2021-08-30.

\bibitem{WinNT4}
``{[Counterbalancing]} counterbalancing to reduce carryover effect.''
  \url{https://opentext.wsu.edu/carriecuttler/chapter/experimental-design/},
  accessed: 2021-08-30.

\bibitem{WinNT6}
``{[Counterbalancing]} counterbalancing to reduce fatigue and practice
  effect.''
  \url{https://psychology.illinoisstate.edu/jccutti/psych231/SP01/week8.html#:~:text=Advantages%3A%20The%20biggest%20advantage%20is%20that%20exposure%20to,levels%20of%20the%20IV%20Counterbalancing%20is%20not%20required},
  accessed: 2021-08-30.

\bibitem{millner2020suicide}
A.~J. Millner, M.~D. Lee, K.~Hoyt, J.~W. Buckholtz, R.~P. Auerbach, and M.~K.
  Nock, ``Are suicide attempters more impulsive than suicide ideators?''
  \emph{General hospital psychiatry}, vol.~63, pp. 103--110, 2020.

\bibitem{plechata2019age}
A.~Plechat{\'a}, V.~Sahula, D.~Fayette, and I.~Fajnerov{\'a}, ``Age-related
  differences with immersive and non-immersive virtual reality in memory
  assessment,'' \emph{Frontiers in psychology}, p. 1330, 2019.

\bibitem{sheskin2018thechildlab}
M.~Sheskin and F.~Keil, ``Thechildlab. com a video chat platform for
  developmental research,'' 2018.

\bibitem{kahraman2018gender}
B.~O. Kahraman, T.~Kahraman, O.~Kalemci, and Y.~S. Sengul, ``Gender differences
  in postural control in people with nonspecific chronic low back pain,''
  \emph{Gait \& Posture}, vol.~64, pp. 147--151, 2018.

\bibitem{faraldo2012influence}
A.~Faraldo-Garc{\'\i}a, S.~Santos-P{\'e}rez, R.~Crujeiras-Casais,
  T.~Labella-Caballero, and A.~Soto-Varela, ``Influence of age and gender in
  the sensory analysis of balance control,'' \emph{European Archives of
  Oto-Rhino-Laryngology}, vol. 269, no.~2, pp. 673--677, 2012.

\bibitem{schedler2019age}
S.~Schedler, R.~Kiss, and T.~Muehlbauer, ``Age and sex differences in human
  balance performance from 6-18 years of age: a systematic review and
  meta-analysis,'' \emph{PLoS one}, vol.~14, no.~4, p. e0214434, 2019.

\end{thebibliography}
\end{document}